\documentclass[longauth]{aa} 
\usepackage{graphicx}
\usepackage[usenames,dvipsnames]{color}
\usepackage{txfonts}
\begin{document}

   \title{The panchromatic spectroscopic evolution of the classical CO  nova V339 Del (Nova Del 2013) until X-ray
turnoff}


   \author{S. N. Shore\inst{1,2}, E. Mason\inst{3}, G. J. Schwarz\inst{4},  F. M. Teyssier\inst{5}, C. Buil\inst{6}, I. De Gennaro Aquino\inst{7},   
K. L. Page\inst{8}, J. P. Osborne\inst{8},  S. Scaringi\inst{9},  S. Starrfield\inst{10}, H. van Winckel\inst{11}, R. E. Williams\inst{12}, C. E. Woodward\inst{13}  
} 

\institute{
Dipartimento di Fisica "Enrico Fermi'', Universit\`a di Pisa;
\and 
INFN- Sezione Pisa, largo B. Pontecorvo 3, I-56127 Pisa, Italy,\\ \email{shore@df.unipi.it} 
\and
Osservatorio Astronomico (INAF) Tireste, Via G.B. Tiepolo 11, I-3413 Trieste, Italy
\and
American Astronomical Society, 2000 Florida Ave., NW, Suite 400, DC 20009-1231, USA
\and
67 rue Jacques Daviel, 76100, Rouen, France
\and
Castanet Tolosan Observatory, 6 place Clemence Isaure, 31320, Castanet Tolosan, France
\and
Hamburger Sternwarte, Gojenbergsweg 112, D-21029 Hamburg, Germany
\and
Department of Physics and Astronomy, University of Leicester, University Road, LE1 7RH Leicester, UK
\and
Instituut voor Sterrenkunde, KU Leuven, Celestijnenlaan 200D, 3001 Leuven, Belgium
\and
School of Earth and Space Exploration, Arizona State University, PO Box 871404, Tempe, AZ 85287-1404
\and
Max-Planck-Institut fuer extraterrestrische Physik, Giessenbachstra$\beta$e 1, 85741 Garching, Germany
\and
Space Telescope Science Institute, 3700 San Martin Drive, Baltimore, MD 21218 USA
\and
 Department of Astronomy ,  University of Minnesota, 116 Church Street S.E., Minneapolis, MN 55455, USA 
}

   \date{Received: ; accepted:}

 
  \abstract
 {Classical novae are the product of thermonuclear runaway-initiated explosions occurring on accreting white dwarfs.  }
   {V339 Del (Nova Delphinus 2013)  was one of the brightest classical novae of the last hundred years.  Spectroscopy and photometry are available from $\gamma$-rays through infrared at stages that have frequently not been well observed.  The complete data set is intended to provide a benchmark for  comparison with modeling  and for understanding  more sparsely monitored historical classical and recurrent  novae.   This paper is the first in the series of reports on the development of the nova.   We report here on the early stages of the outburst, through the X-ray active stage.}
{ A time sequence of optical, flux calibrated high resolution spectra was obtained with the Nordic Optical Telescope (NOT) using FIES  simultaneously, or contemporaneously, with  the Space Telescope Imaging Spectrograph (STIS) aboard the Hubble Space Telescope  during the early stages of the outburst.  These were supplemented with  MERCATOR/HERMES optical spectra.  High resolution IUE  ultraviolet spectra of OS And 1986, taken during the Fe curtain phase, served as a template for the distance determination.  We used standard plasma diagnostics (e.g., [O III] and [N II] line ratios, and the H$\beta$ line flux) to constrain electron densities and temperatures of the ejecta.  Using Monte Carlo modeling of the ejecta, we derived the structure, filling factor, and mass from comparisons of the optical and ultraviolet line profiles. \footnote{
Based on observations made with the NASA/ESA Hubble Space Telescope, obtained [from the Data Archive] at the Space Telescope Science Institute, which is operated by the Association of Universities for Research in Astronomy, Inc., under NASA contract NAS 5-26555. These observations are associated with program \# 13828.

}
  }
{We derive an extinction of E(B-V)=0.23$\pm$0.05 from the spectral energy distribution,  the interstellar absorption, and H I emission lines.   The distance, about 4-4.5 kpc, is in agreement with the inferred distance from near infrared interferometry.    The maximum velocity was  about 2500 km s$^{-1}$, measured  from the UV resonance and optical  profiles.  The ejecta showed considerable fine structure in all transitions, much of which persisted as emission knots.  The line profiles  were modeled using a bipolar conical  structure  for the ejecta within a relatively restricted range of parameters.  For V339 Del, we find that an inclination to the line of sight  of about $35^o - 55^o$, an opening angle of $60^o-80^o$, and an inner radius $\Delta R/R(t)\approx 0.3$ based on $v_{\rm rad, max}$ matches the permitted and intercombination lines.  The filling factor is $f \approx$ 0.1, and the derived range in the ejecta mass is $ (2-3)\times 10^{-5}$M$_\odot$.  
  }
  { }

 
 \authorrunning{Shore et al.}
 \titlerunning{V339 Del (Nova Del 2013) during its early evolution}

   \maketitle
%

\section{Introduction}

Nova Del 2013, designated V339 Del, was discovered at unfiltered CCD magnitude 
6.8  by  Itagaki (AAN  2013)  on  2915 Aug 14.58 UT, and spectroscopic
confirmation  was provided shortly afterward  (14.92 UT)  by  Masi (CBET
3628).  The  nova was  discovered  during  its rising  phase.  Optical 
maximum was on  Aug. 16.25  UT when it  reached V$\sim$4.3 mag  (Munari et
al. 2015).  It  rapidly became a popular target  among amateurs  (e.g., ARAS, AAVSO) and  professional astronomers because of its brightness.
It has been detected in the range $\gamma$-rays (Ackermann et al. 2014) through centimeter radio (ATel 5298, 5376).  A study of the carbon spectrum in V339 del was published in Shore et al. (2014).  A high resolution atlas of the first four months of the outburst, covering 3800-8800\AA, has been published by De Gennaro Aquino et al. (2015) and a discussion of the infrared development by Gehrz et al. (2015).

We are spectroscopically monitoring V339 Del at high resolution starting on Day 13 as  part of  the international  campaign  using coordinated optical  (NOT/FIES) and UV (HST/STIS) observations.   Our intent is to produce a benchmark data set of calibrated, high resolution panchromatic observations, covering the range from 1150 - 7400\AA, that would serve for modelers and as a baseline for comparison with less well covered  historical classical and recurrent novae.   By observing at specific epochs in the UV and optical bands,  we sampled all  the important resonance and excited state transitions of the key elements (especially H, He, C, N, O, and Ne) up through the four times ionized state.   Our  goal  is to spectroscopically determine the  density, velocity, and spatial distribution for different elements within the ejecta and how these properties change over time.  We use these data to obtain information about  the ejecta geometry,  ionization structure, filling factor, and  mass.  Issues of chemical 
composition and homogeneity will be addressed in a future paper.  In this first report,  we present  an analysis of the early decline and the X-ray supersoft source  (SSS) plateau stage.  The temporal development of the ejecta ionization and density structure, and comparisons with other recent novae similarly analyzed, are postponed to the next paper once our late nebular  spectra have been taken.    

In this paper, we check and extend the conclusions drawn from our earlier time development study of the recurrent nova T Pyx (Shore 2013, De Gennaro Aquino et al. 2014).   T Pyx showed that the spectra arise in ballistically expanding,  bipolar, highly fragmented ejecta.   The velocity and density structures were modeled by simple radial power laws resulting from a ballistic expansion and there was no indication of a continuous outflow (i.e. winds) or multiple ejections.   These structures were frozen in the expanding gas and evolved spectrally because of  the changing illumination and continuum of the  cooling white dwarf.  Thus, the observed structures point back to the moment of origin during the TNR and the advent of the explosion.  We show in this paper that the same conclusions hold for a CO nova. 


\section{Observations and data reduction}

 \begin{table}
\flushleft
 \caption[]{Log of observations. \textcolor{blue}{}}
 \label{log}
\scriptsize
\begin{tabular}{lccccc}
obs-date (start) & age & epoch & instrument & setup & exptimes  \\
		 & (days) &    &            &       & (s)   \\
2013-08-28.945   & 14  & Fe-curtain & NOT/FIES & HiRes  & 100+1000+10  \\
2013-09-08.006   & 25  & Fe-curtain & NOT/FIES & HiRes  & 30+1000 \\
2013-09-18.205   & 35  & Fe-curtain & HST/STIS & E140M/1425 & 108 \\
2013-09-18.210   & 35  & Fe-curtain & HST/STIS & E230H/1763 & 105 \\
2013-09-18.216   & 35  & Fe-curtain & HST/STIS & E230H/2013 & 105 \\
2013-09-18.221   & 35  & Fe-curtain & HST/STIS & E230H/2263 & 105 \\
2013-09-18.226   & 35  & Fe-curtain & HST/STIS & E230H/2513 & 105 \\
2013-09-18.231   & 35  & Fe-curtain & HST/STIS & E230H/2762 & 105 \\
2013-09-30.858   & 46  & Fe-curtain & NOT/FIES & HiRes  & 2000+300+30 \\
2013-10-27.892	 & 73  & transition & NOT/FIES & HiRes  & 300+3000 \\
2013-11-21.831	 & 99  & transition & NOT/FIES & HiRes  & 3000 \\
2013-11-21.916   & 99  & transition & HST/STIS & E140M/1425 & 575 \\
2013-11-21.926   & 99  & transition & HST/STIS & E230M/1978 & 575 \\
2013-11-21.937   & 99  & transition & HST/STIS & E230M/2707 & 575 \\
2013-12-15.809	 & 122 & transition & NOT/FIES & HiRes  & 3000 \\
2014-04-15.125	 & 243 & late-SSS & NOT/FIES & HiRes  & 3600 \\
2014-04-19. 021  & 248  & late-SSS & HST/STIS & E140M/1425 & 575 \\
2014-04-19. 070  & 248 & late-SSS & HST/STIS & E230M/1978 & 575 \\
2014-04-19.081   & 248 & late-SSS  & HST/STIS & E230M/2707 & 575 \\
2014-04-19. 086  & 248 & late-SSS & HST/STIS & G430L & 10 \\
2014-04-19. 090  & 248 & late-SSS  & HST/STIS & G430M/3423 & 50 \\
\end{tabular}
\end{table}

The data presented in this paper were primarily collected  with
HST/STIS (program 13388) and  NOT/FIES (program 48-103).  Table 1 gives the log of observations.  
The HST/STIS program consisted in a
non-disruptive  Target of Opportunity (ToO) for the early observation  of bright Galactic  novae. Each visit was just one orbit
long. The instrument setup was chosen to cover the whole UV range from 1200 to 2900 \AA. In addition, the choice of the echelle resolutions  (E230H or E230M) was driven by the target  magnitude, as was the aperture (0.2$\times$0.2 arcsec  in all cases), satisfying the requirements of high-spectral-resolution and instrument safety constraints.  Each HST/STIS observation was triggered on the basis of the SWIFT UV magnitude obtained from UVOT and X-ray evolution complemented with the optical light curves, thus to cover the most important evolutionary phases of the ejecta.   On 2014 Apr. 19 we also obtained HST/STIS low (G430L) and a medium (G430M) resolution spectra covering the wavelength ranges 2900-5700 \AA\ and 3290-3560 \AA, respectively. These were intended to cover  spectral intervals not otherwise obtained with the echelle spectra and to provide an additional check on the flux calibration.   
The ground observations were activated in parallel to HST ToO and scheduled on the closest interval available for queue/service mode operations.   Several NOT observations were  taken as  a stand-alone  between combined STIS+NOT epochs and serve as a bridge between epochs.

The early decline of the nova was independently observed by MERCATOR/HERMES 
 ($R \approx 86000$) (see Raskin et al., 2011) which provided monitoring with an almost daily cadence during the first 100 days after discovery. Of these data we present a selection from the first 54 days of monitoring that span epochs not otherwise covered by our STIS+FIES programs and were useful for the kinematic analysis of the ejecta
and its substructures along the line of sight during the optically thick phase. 

HST/STIS spectra were reduced with the MAST online pipeline. The spectral orders where then merged with the IDL
routine originally developed for GHRS by the HST team (CALHRS).   The FIES spectra were preprocessed, extracted, and
merged with the FIES pipeline. The spectra were subsequently flux calibrated with custom routines written in IDL and  IRAF.  We made sure that each FIES observation of the nova was accompanied by the observation of a spectrophotometric standard star, BD +28$^o$4211,  at similar airmass and pointing direction to minimize the color-dependent losses typical of fiber-fed spectrographs.  We verified  that the uncertainties in our flux calibration are at most 20\%\ by comparison with photometric observations and lower resolution spectra (including from space). When necessary, the  optical spectrum was offset in flux to match the level of the STIS UV spectrum.    
HERMES spectra were also preprocessed, extracted and merged by the instrument pipeline. However, they are not flux
calibrated, since no spectrophotometric standard star was observed, and the spectrograph signatures  (e.g., residual ripple) were not fully removed.

\section{X-ray observations with {\it Swift}}

  \begin{figure}
   \centering
   \includegraphics[width=6.5cm,angle=-90]{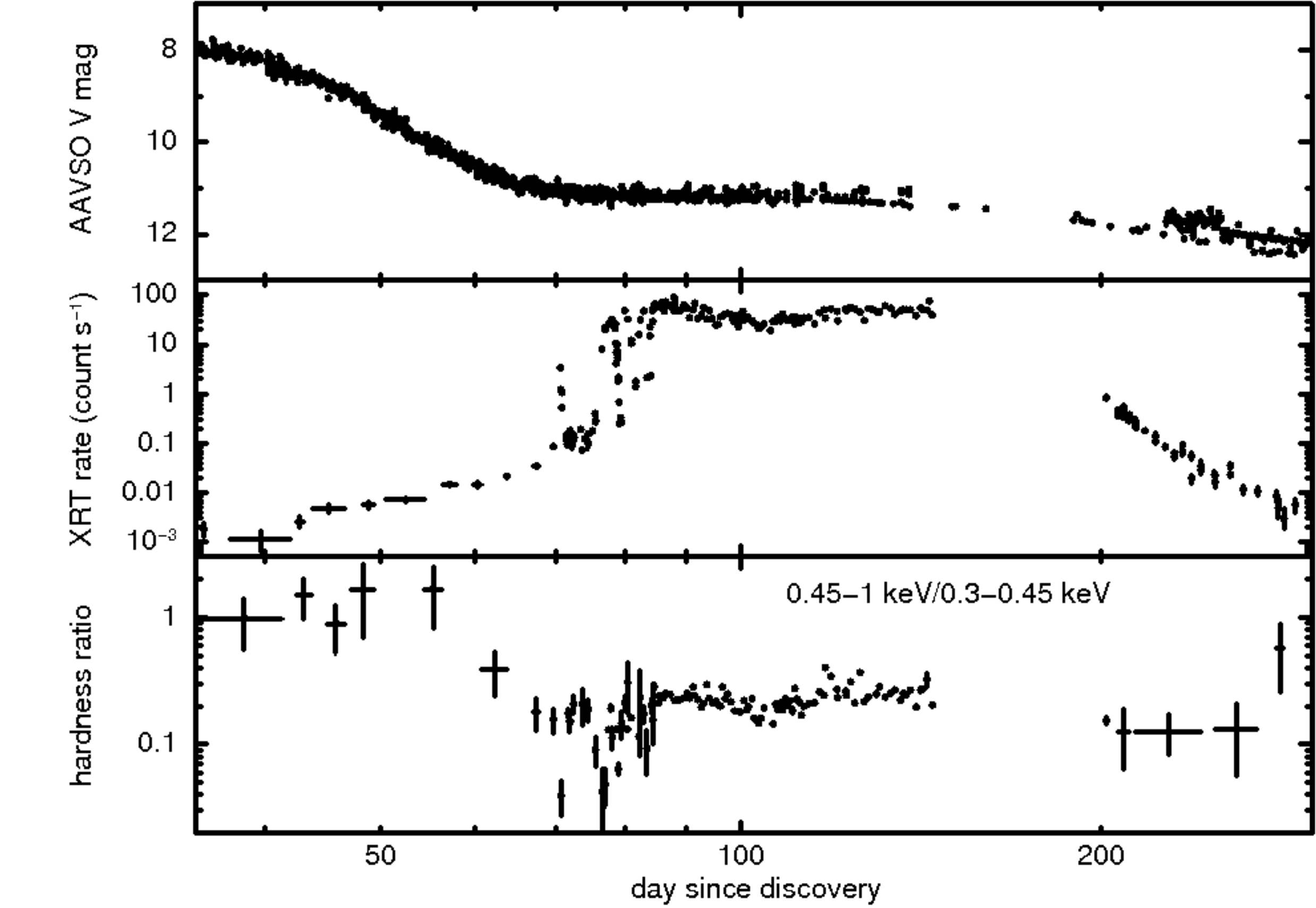}
      \caption{Top:  AAVSO visual light curve.  Middle: {\it Swift} XRT count rate for the period covered by this paper, through Day 300.  Bottom: Hardness ratio F(0.45-1 keV)/F(0.3-0.45 keV).  The SSS phase is defined to be when the HR$<$1.  The first FIES/STIS epoch was before, and the second after the appearance of the SSS.}
         \label{f0}
   \end{figure}

   \begin{figure*}[ht]
   \centering
   \includegraphics[width=14cm,angle=0]{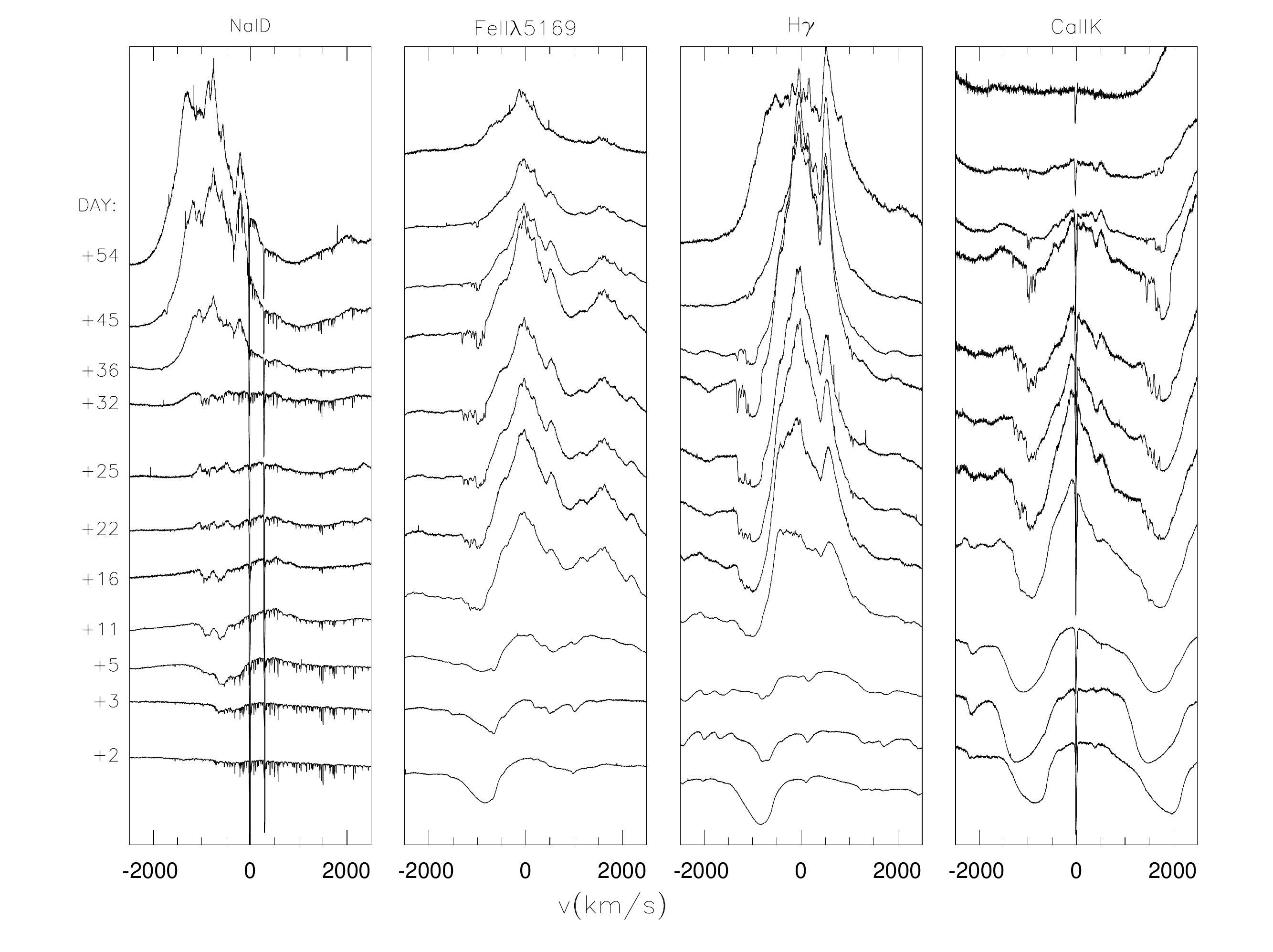}
   \caption{Line profile evolution of the transitions NaI D, Fe II 5169\AA, H$\gamma$, and Ca II K in the early
decline (day +2 to day +54) HERMES spectra. In each panel time increases from bottom to top. The epoch of each spectrum
being on the left side of the figure. The y axis unit (ADU) have been omitted since the spectra are not flux calibrated
and the spectral signatures have not been properly removed. 
However, day 2 transitions are plotted as in the original spectrum with the 0 counts level matching the x axis. The
subsequent epochs have been offset by 0.75 ADUs in the case of the NaI D, 0.70 ADUs in the case of the Fe II 5169\AA\ 
transition, 0.85 ADUs in the case of H$\gamma$ and 3.5 ADUs in the case of the CaII K transition. 
}
              \label{f1}
    \end{figure*}
    
We include a brief description of the nova's X-ray development during our period of observation, since the central white dwarf is the photoionization source that drives the spectroscopic evolution of the ejecta in the optical and ultraviolet  (cf. Woudt \& Ribeiro 2014).   {\it Swift} observations of V339 Del began on 2013 Aug 14.96 UT.   Starting at only nine
hours after its discovery, the follow-up for V339 Del was one of the most rapid responses   
 to a nova performed by {\it Swift}. Observations were obtained 
regularly until the nova entered the Sun constraint for {\it Swift} at 144.5 days
after discovery. Data collection resumed on day 202. 

A weak X-ray source was first detected by the XRT (X-ray Telescope) (see Burrows et al. 2005) 35.6 days after the
nova was  discovered. The source then brightened by almost five
orders of magnitude, passing through a period of large-amplitude flux
variability before reaching a peak count rate of $\sim$90 count s$^{-1}$ (0.3-10 keV) around
day 88. The X-ray source then remained bright, with a slow
variation between $\sim$80 and 20 count\ s$^{-1}$, until the nova was too close
to the Sun for observations. After re-emerging from the constrained zone, it was detected 
at  about 1 count s$^{-1}$.  There followed a monotonic fading until observations ceased
on day 373.  Figure 1 shows the AAVSO V light curve (top panel), 0.3-10 keV X-ray light curve (middle panel), and the
hardness ratio (HR), defined as the ratio of the counts in the 0.45-1.0 keV and
0.3-0.45 keV bands (bottom panel). The bright X-ray source was {\it extremely}  
soft, with very few counts being detected above 1 keV, hence our choice of energy
bands.

   \begin{figure*}[ht]
   \centering
   \includegraphics[width=14cm,angle=0]{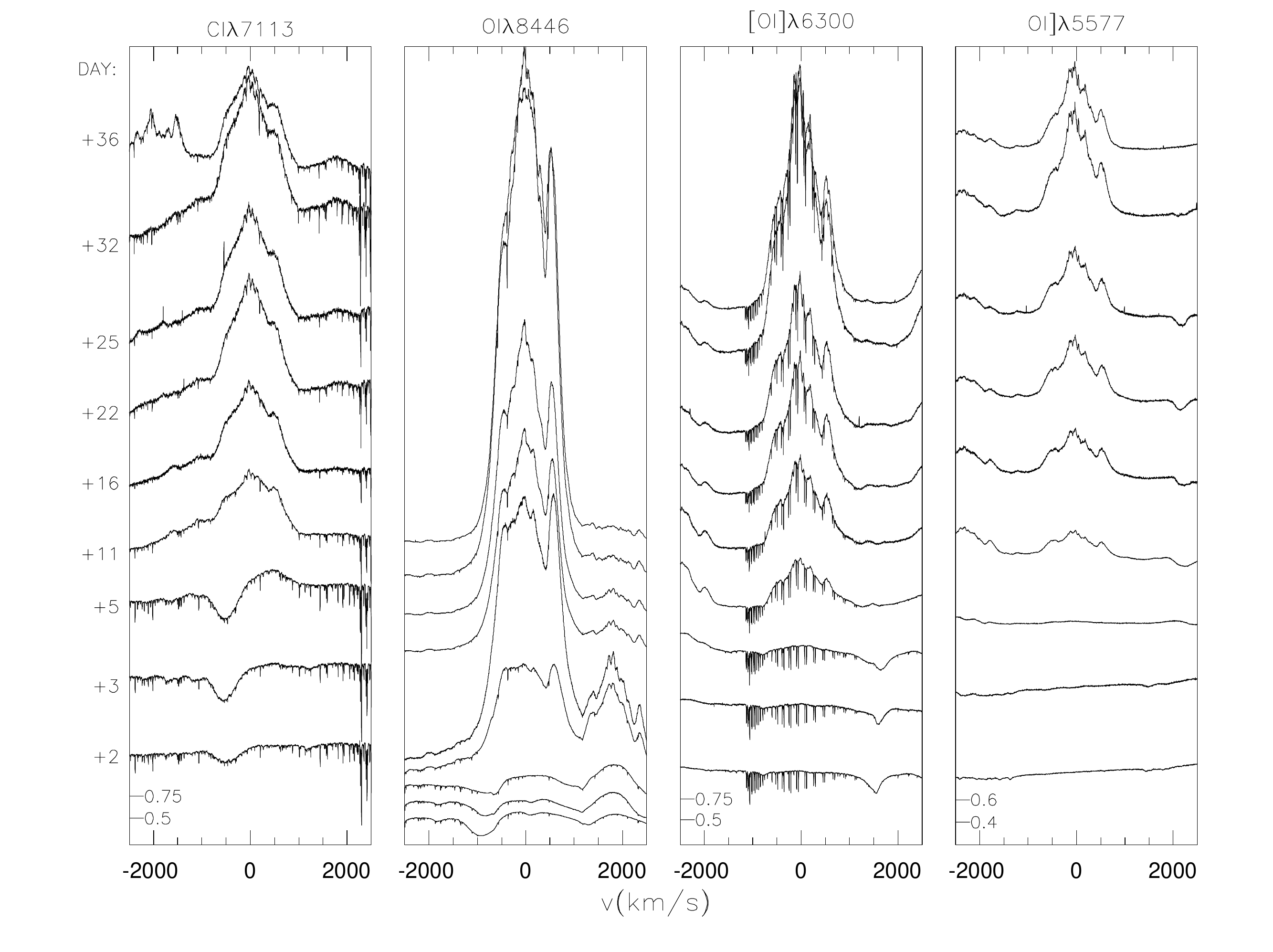}
   \caption{Line profile evolution of neutral C and O during the early decline monitored at MERCATOR/HERMES. The
C I 7713\AA, [O I] 5577\AA, and [O I ] 6300\AA\ lines are offset with respect to each other by 0.8, 0.75 and
0.75, respectively. The first 5 spectra of OI  8446\AA\ were offset by 0.5 while the last four were scaled to 0.25
and offset by 1.33$\times n$, with $n$ being the spectrum position in the sequence. The ``0.5,0.75'' and
``0.4-0.6'' tick-marks indicate the scale in those panels whose x-axis is not matching the 0 ADU level of the y-axis.
}
              \label{f2}%
    \end{figure*}

\section{Contemporary ultraviolet and optical spectroscopic development from Day 35 to Day 248}

Our coordinated ground and HST/STIS  high resolution spectroscopic observations cover three fundamental stages of the
ejecta and white dwarf evolution: i) the iron-curtain (day +35 STIS spectrum  with the bracketing FIES observations
of day +25 and +46); ii) the transition phase (day +99 STIS and 
FIES spectra); and iii) the nebular phase after the peak of the X-ray super-soft source (SSS) phase.  The nuclear source remained active on the white dwarf during this interval (the X-ray spectrum was still soft) but its luminosity was decreasing (day 243 FIES spectrum and day 248 STIS spectrum).

The optically thick stage (hereafter called the ``Fe curtain'') is especially difficult to interpret in the  UV since the complex overlapping bands of absorption lines from low ionization metallic species produces a strongly wavelength dependent pseudo-continuum.     The spectral distribution also contains a number of ``windows'', regions of reduced line opacity not associated with any strong transition but, instead, arising from photons escaping from the deeper, hotter  parts of the ejecta (e.g., Hauschildt et al. 1992).  The appearance (intensity, wavelength) of the pseudo-emission features depends on the velocity and density structure of the ejecta and the luminosity of the underlying white dwarf.   The emission in any spectral interval depends on the velocity gradient, which for  ballistic ejection is a constant that depends only on the maximum expansion velocity.  In spectral regions where the optical depth is very large, and the radius of the effective photosphere is nearly that of the outermost 
material (e.g. far UV), there is no emission 
and virtually no flux.  In contrast, in regions of lower opacity (near UV and optical), the volume filled by low optical
depth gas is proportionally higher, and emission lines may occur.  This is illustrated by the first STIS spectrum of
V339 Del.  Shortward of 2000\AA, there were few genuine emission lines, the O I 1302,1304\AA\ and C II 1334,1335\AA\
doublets being likely exceptions (Fig. A1).  Longward, several emission lines, notably N II 2143\AA, C II 2423\AA\,
and Mg II 2796,2803\AA\, were clearly present (Fig. A1).   No other absorption feature can be identified at this
stage with any certainty since the line formation is highly stratified and depends on the detailed coupling of
individual transitions that overlap and shadow each other (see, e.g. Hauschildt et al. 2002, Shore 2012).


The optical range (NOT/FIES spectra from day 13, 24 and 46, and MERCATOR/HERMES spectra from day 2 to day 45), having lower optical depth than the UV, was characterized by numerous emission lines from Fe II and low ionization-energy ions (Fig. A1).  At this time, we observed  distinct velocity ranges of the absorption components and a characteristic emission component common to all  transitions of similarly small optical depth.   The HERMES spectral  sequence in Fig. 2 shows that during the first  five days after discovery the lines were dominated by P Cyg-like profiles, with the absorption trough extending to high velocities:  -1500 km s$^{-1}$ (H$\gamma$), -1600 km s$^{-1}$ (FeII and CaII). The NaI D lines did not appear until day 3, and its absorption trough never extended beyond  -1000 km s$^{-1}$.   Short lived (day 2 to 5) low velocity absorptions were also observed in CI and OI\footnote{For O I 7773\AA, not shown in the figure, the absorption components were present until day 11 decreasing from -1000 
to -600 km s$^{-1}$ while weakening; see De Gennaro Aquino et al. (2015) and the discussion in Shore et al. (2014).}, for which we measure $v_{rad} \sim$ -500 and -1000 km s$^{-1}$, respectively (see Fig. 2). At this early stage, the absorption troughs were featureless and unsaturated. 

\begin{figure}
   \centering
   \includegraphics[width=9cm]{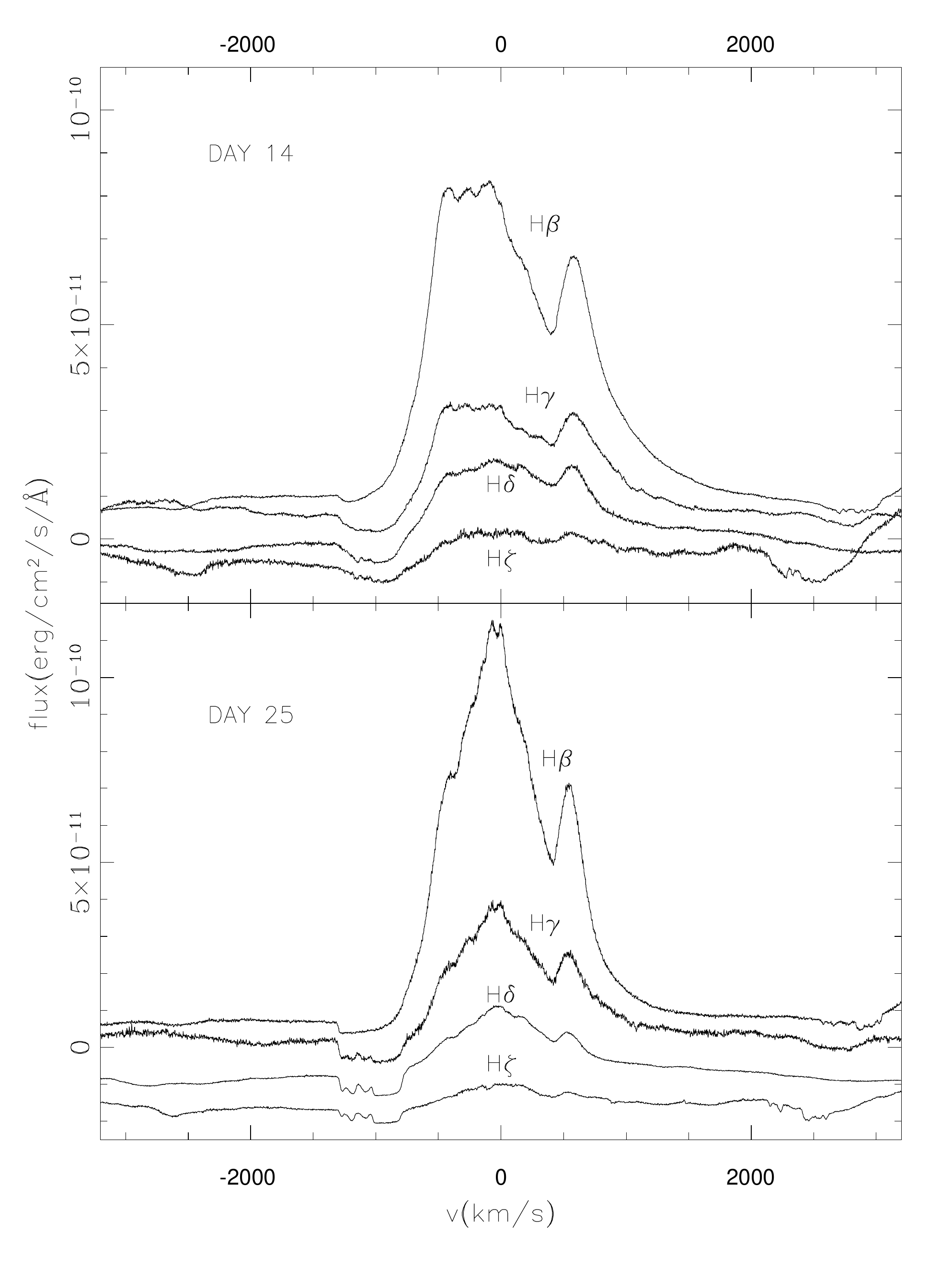}
      \caption{Line profile of the hydrogen Balmer lines H$\beta$, H$\gamma$, H$\delta$, H$\zeta$ at day +14 (top panel) and
+25 (bottom panel) in the NOT/FIES spectra. The spectra were flux calibrated and the H$\gamma$, H$\delta$, H$\zeta$
lines have been offset with respect to H$\beta$ by the constants -0.5e-11, -1e-11 and  -1.5e-11, respectively, in the
+14 days spectrum and by the constants -0.75e-11 -1.5e-11 and -2.25e-11, again respectively, in the +25 days spectrum.  Note the Balmer progression in the absorption trough: decreasing strength and velocity of the feature and increasingly fragmented  structure  with increasing series number.
              }
         \label{f3}
   \end{figure}

From day 11 the absorption trough fragmented into narrow substructures, while the maximum radial velocity of the ensemble of 
absorption components reduced to $\sim$-1350 to -1400 km s$^{-1}$ for FeII and CaII and H (the bulk of the absorption being
centered at $\sim$ -1000 km/s). The NaI D lines remained at lower velocity, centered at $\sim$ -900  km s$^{-1}$.  
By the time the absorption substructures disappeared (at end of the Fe curtain, day 54),  the emission lines had developed broad wings that were at least as extended as the maximum velocity of the absorptions components of the day 2 and day 3 spectra (H and Fe).  The hydrogen absorption troughs followed the velocity decrement caused by the decreasing optical depth along the line series.  The maximum negative velocity was, however, the same across the whole Balmer series and reached about -1300 to -1400 km/s (Fig.\ref{f3}).

The P Cyg {\it emission} components, when present, were featureless as were the absorptions (see Figs.\ref{f1} and \ref{f2}).  The emission  lines grew progressively stronger and more complex with time.  The hydrogen Balmer and neutral O permitted lines were initially flat-topped, caused by the high line optical depth (day $>$5 to 16) and, later, both developed a more ``peaked''
profile (Figs.\ref{f1},\ref{f2} and \ref{f3}).  The profiles were soon alike for all lines, with a
distinct emission peak at about +600 km s$^{-1}$, although with no corresponding component at -600 km
s$^{-1}$, and displaying a substantial, complex emission in the velocity interval $\left| v_{rad}\right| <$ 500 km s$^{-1}$.  This 
profile asymmetry can only in part  be attributed  to an opacity effect (for example, Fig.\ref{f3}), since it was also present on the forbidden transitions [O I ] 5577, 6300\AA\ (Fig.\ref{f2}).  

 \begin{figure}
   \centering
   \includegraphics[width=9cm]{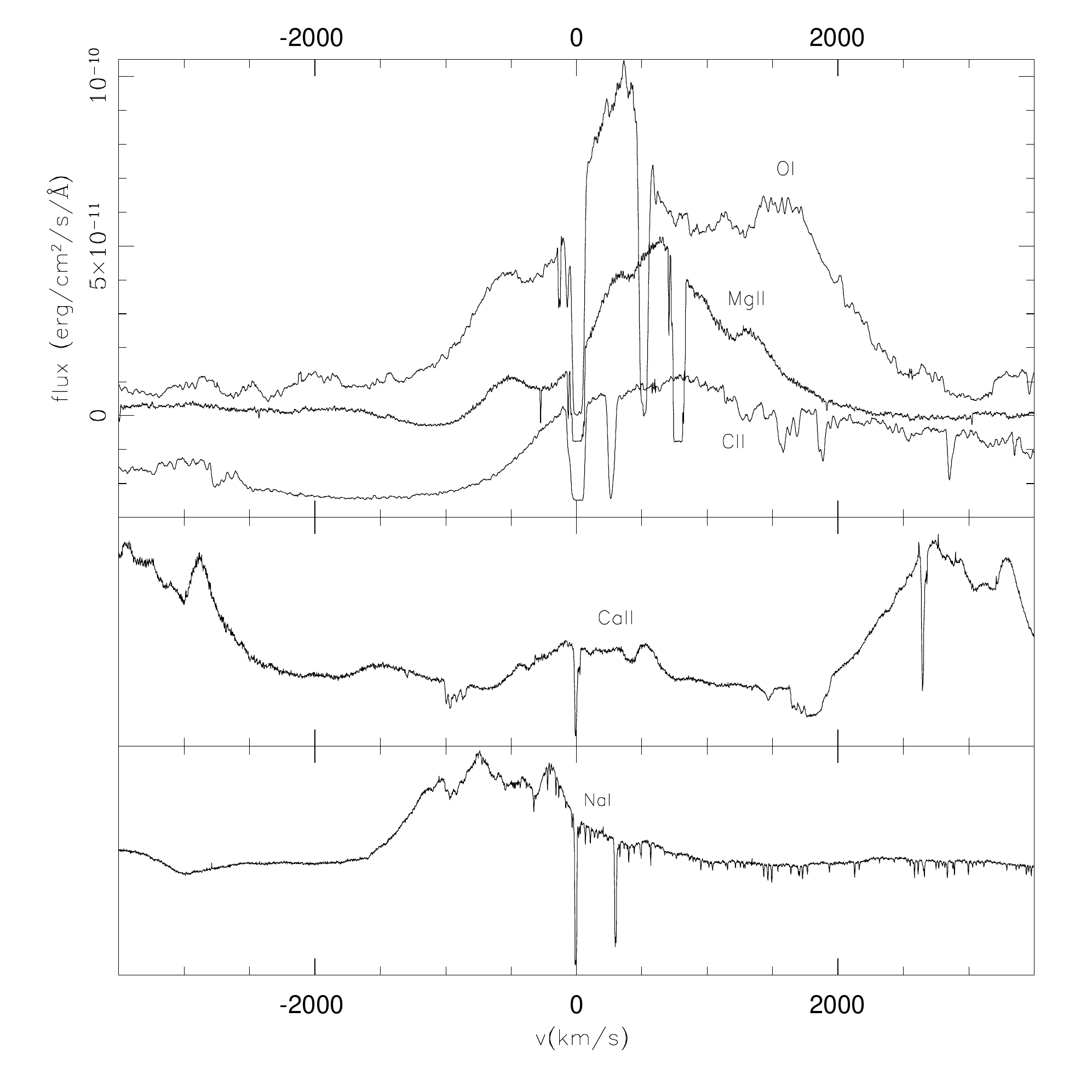}
      \caption{Comparison of the line profiles of the UV resonant lines (MgII 2800\AA, CII 1334\AA, 
and OI 1302\AA) from day 99 with the optical ones NaI D and CaII K (CaII H, in the figure, is blended with H$\varepsilon$).
The NaI D had almost disappeared and was blended with He I 5876\AA\ emission which was the strongest line in the 5800-6000\AA\ range  by this time. 
              }
         \label{f4}
   \end{figure}


The intensity of the Fe II emission began decreasing around day 33, when the He I emission lines (e.g., 5876\AA\
and 7065\AA) first appeared.  We take this as a  signature that the UV  Fe curtain had started its gradual lifting and that
the ejecta were patchy. At UV wavelengths, our first HST/STIS observation (day 35) displayed a still heavily veiled
continuum, although resonance lines, such as Mg II 2800\AA, O I 1302\AA, and C II 1334,1335\AA\, were  visible
(Figs. A1 and 5). We do not observe absorption components except for the Mg II doublet and
the redder component of the O I doublet.  What appears to be an  absorption trough of the C II doublet ,extending to 
-2500 km s$^{-1}$, is probably spurious, since the blue wing of the emission remains the same during the subsequent epochs. 

   \begin{figure*}[ht]
   \centering
   \includegraphics[width=13cm,angle=-90]{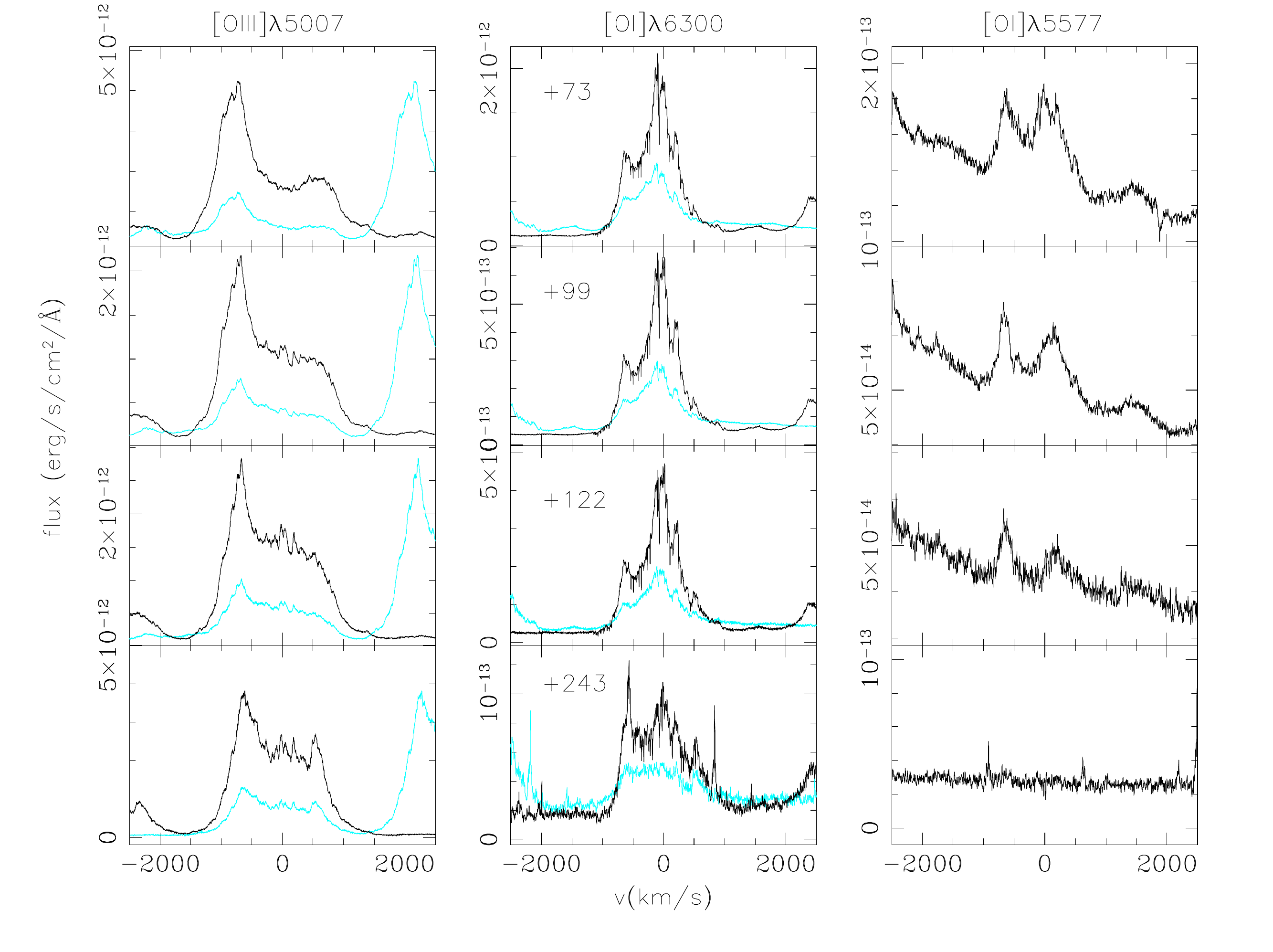}
      \caption{O I and O III evolution from the NOT sequence.  Left: [O III] 5007\AA\ (black), 4959\AA\ (blue); middle: [O I] 6300\AA\ (black), 6364\AA\ (blue); right: [O I] 5577\AA.}

         \label{f6}
   \end{figure*}

The day 99  combined UV and optical spectrum was marked by signatures of the so-called ``transition phase'':  the appearance
of He II 1640\AA\ and 4686\AA\ emission  and the nebular transitions [O III] 4959,5007\AA\ and 4363\AA\ together
with their isoelectronic counterparts [N II] 6584,6548\AA\ and 5755\AA.  This also coincided with the start of
the  {\it supersoft source} (SSS) stage.   A comparison of some of the resonance line profiles from the UV and optical is shown in Fig., 5.  
The ionization level of the transitions in the UV range had significantly increased as shown by the presence of the N IV] 1486\AA, NV 1238,1242\AA\ and C III] 1909\AA, C IV1548,1550\AA\ (Fig. A2). The  He II 4686\AA\ and the [O III] and [N II] nebular lines were already present in our day 73 NOT/FIES spectrum,  implying by the higher ionization and photo-excitation that the SSS phase had already begun.  This agrees with the X-ray light curve and hardness ratio variation.   The main difference between the two optical spectra, other than the overall weakening in flux because of the expansion of the ejecta, was that the asymmetry of the Balmer and forbidden line profiles decreased from day 73 to 99. The asymmetry consisted of a pronounced blue peak at $v_{rad} \approx$ -900 km/s substantially broadening the line profile observed during early decline. On the red side of the profile, the lines appeared  unchanged with respect to the early decline, and  the peak at  around +600 km/s (Fig.\ref{f6}) was still 
visible.  

\begin{figure}[h]
   \centering
   \includegraphics[width=9cm]{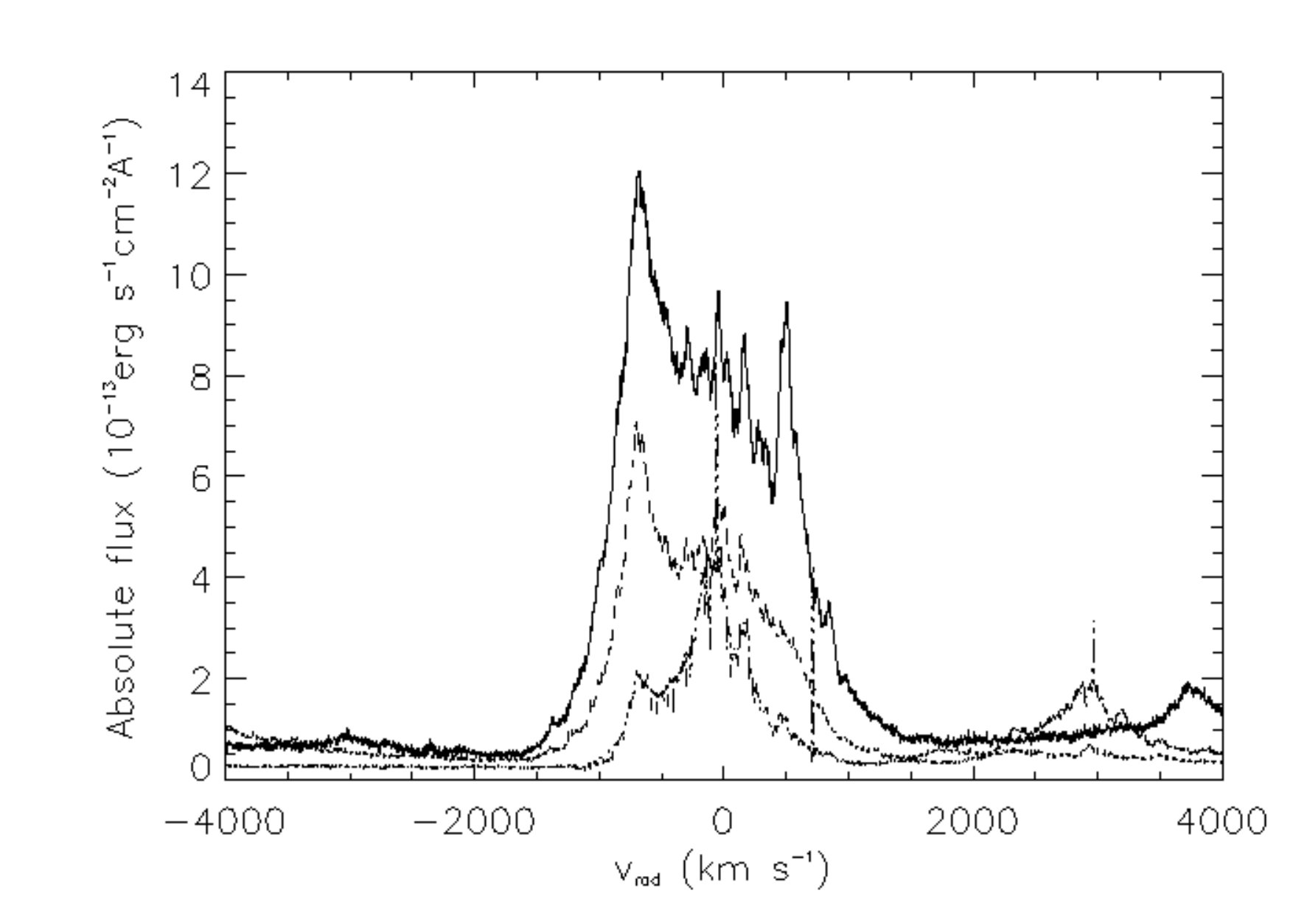}
      \caption{Day 122, for which only optical observations were obtained.  Comparison of the H$\beta$ (solid), [N II]
5755\AA\ (dash), and [O I] 6300\AA\ (dot-dash) line profiles.  No scaling has been applied.  Note the change in the
relative strength of the individual emission features with ionization stage.   }
         \label{f5new}
   \end{figure}

With the ingress to the transition phase,  [O I] 5577\AA, among the oxygen lines, showed the largest change (see Fig. 6). 
While the [O I] 6300\AA\ line profile remained  invariant, 5577\AA\ almost vanished.  Although residual
emission was still present on 5577\AA\ at -680 and +170 km s$^{-1}$, the central emission peak of the profile was absent. The two lines have different origins: [O I] 6300\AA\ can be collisionally excited directly from the
ground state.   It is also fed by the 1641\AA\ line, whose upper state is populated by absorption in the 1304\AA\
resonance transition. In contrast, 5577\AA\ is fed from the intercombination 2324\AA\ transition and has a much lower
branching ratio from the 3$^3$S$^o_1$ state than 1641\AA\ (see Fig. 1 in Shore \& Wahlgren (2010)).   By Day 99, we know
that O I 1304\AA\ emission was extremely weak and O$^o$ was a minor species, but the O I 1025\AA\ line could also be
pumped by Ly$\beta$ throughout the ejecta, leading to 6300\AA\ emission. This may explain the puzzling 
persistence of the neutral oxygen forbidden emission during the transition and early nebular stage without  needing to
invoke extremely opaque knots within the ejecta.

We have only a NOT/FIES spectrum for day 122, but we mention it here  since this was the last obtained before the source was solar constrained from the ground (Fig. 7).   Relative to Day 99, although small changes were observed in individual features, the
lines merely decreased in intensity.  

On Day 248, the combined  UV and optical spectrum (Fig. A3) was similar to that on the day 99 for both observed ionization stages and transitions and species. The lines  just decreased in flux as the gas expands.  Since the profile fine structure may be produced by  density differences among the individual knots, we note that there were only small differential changes between the two dates  (Fig. 9).   We can use this to address a simple question: do the substructures  follow the overall density decrease of the ejecta with time ($t^{-3}$).  From the persistence of the knots and their relative visibilities during the late stages, it seems the answer is ``yes''.  The knots are the ``tip if the iceberg'', the largest fluctuations in the density distribution of the fragments.  Their densities change along with the rest of the gas (see sec. 5.4).  The substructures appear to be density enhancements that are otherwise stationary with respect to the diffuse gas of the ejecta.

   \begin{figure*}
   \centering
   \includegraphics[width=12cm,angle=270]{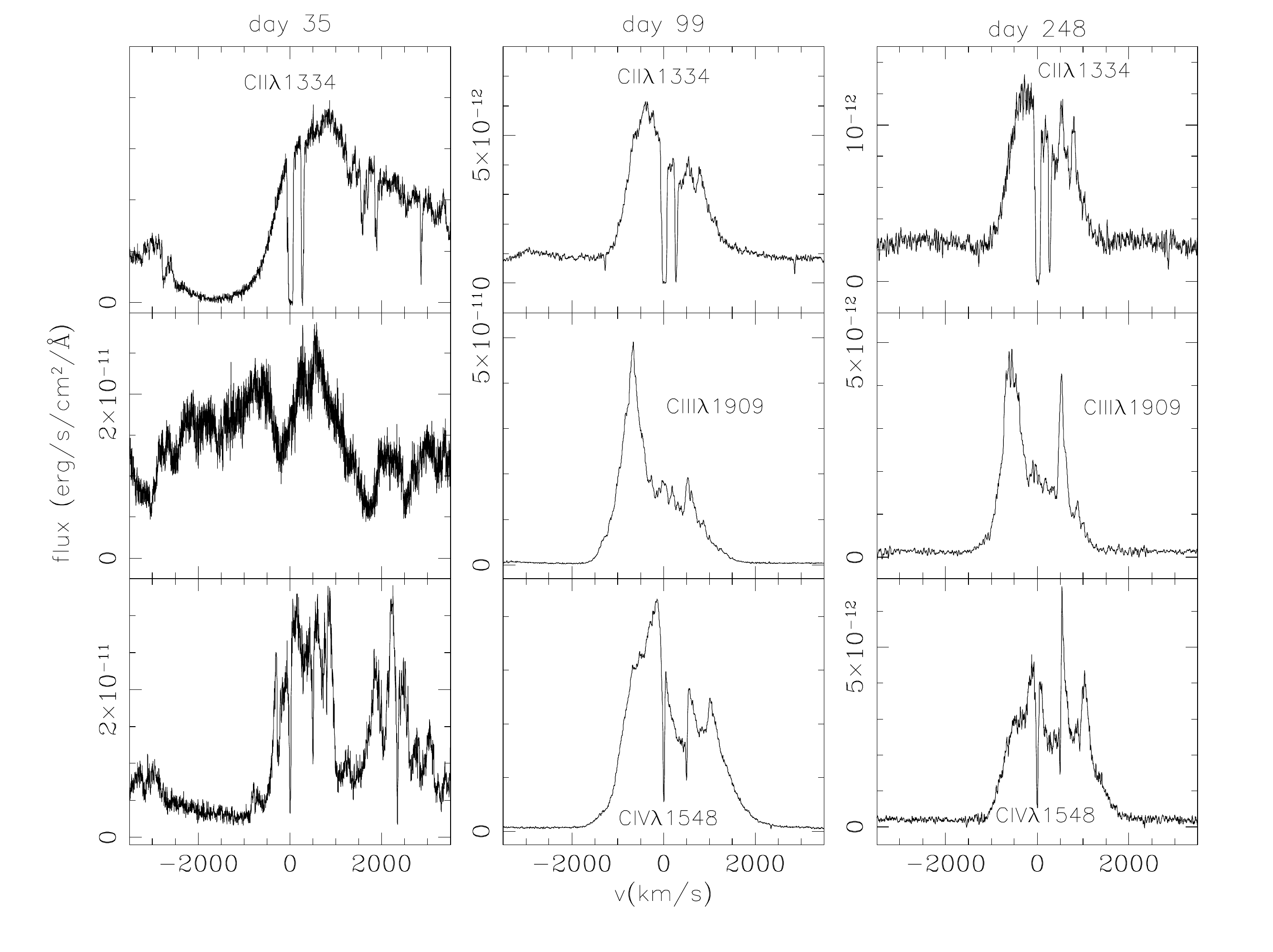}
   \includegraphics[width=12cm,angle=270]{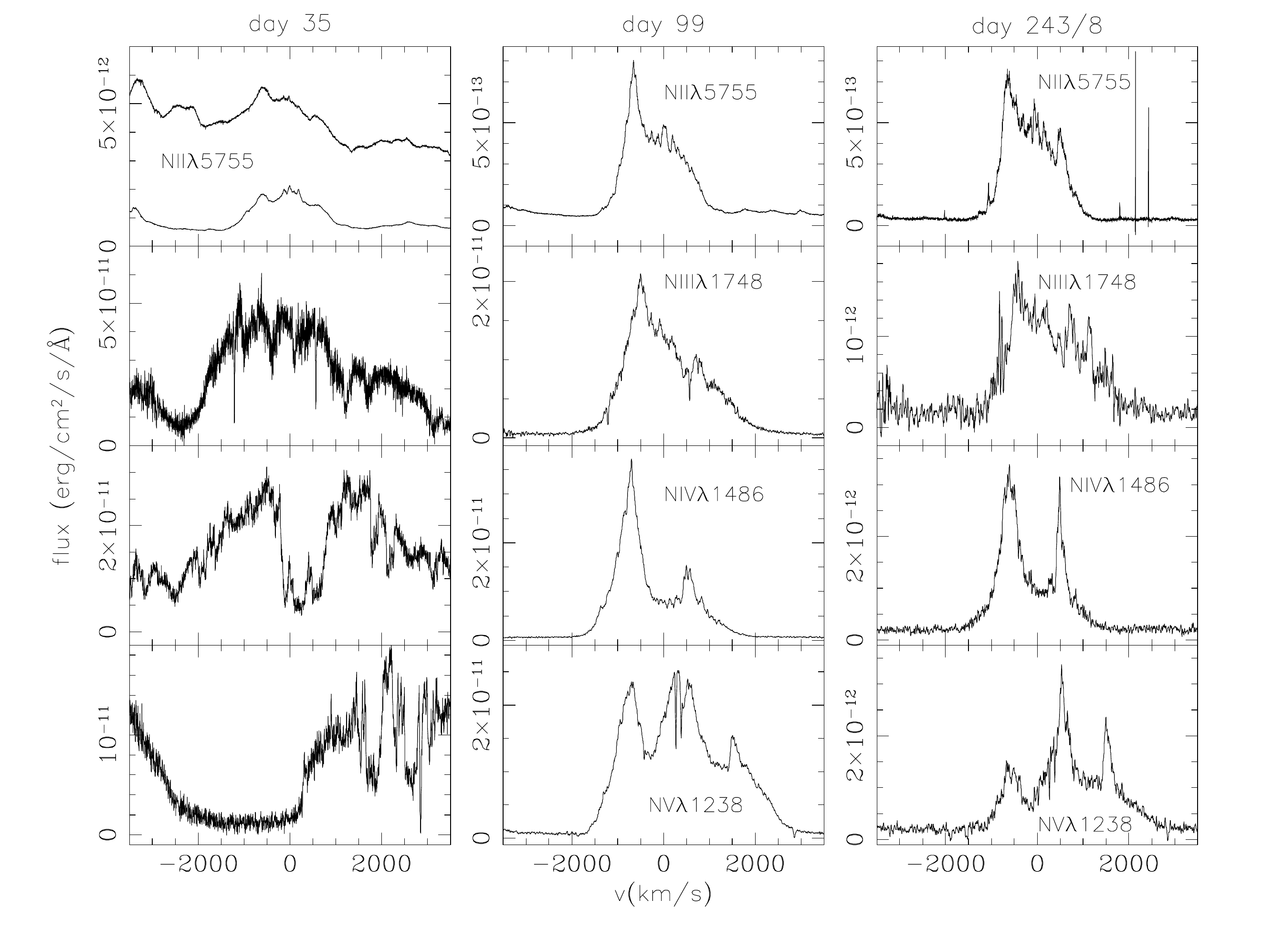}
   \caption{Line profile of the C II-IV and N II-V resonant transition at the various epoch for which we have
contemporaneous optical and UV observations. See text for more details. }
              \label{f5new}%
    \end{figure*}

\section{Discussion}

This panchromatic series of spectra provides a survey of the ionization structure and its variation with time. Four 
ionization stages were available for C (C$^o$-C$^{+3}$) (C I profiles from the earliest spectra are presented in Shore et al. (2014) and de Gennaro Aquino et al. (2015)) and  
N (N$^+$-N$^{+4}$) during the epochs of simultaneous FIES and STIS coverage. This spectral development is
shown in Figs. 8.  During the opaque phase (day 35 STIS spectrum) only C II was visible with a featureless emission
component against which interstellar and Fe curtain absorption components can be distinguished. The C III] and C IV
lines were not detected. The complex ensemble of features visible in the velocity range -3000 to +3000 km/s around each
of them was entirely  caused by the metallic absorption bands. Similarly only [N II] 5755\AA, in the optical, appeared
in emission.  Lacking simultaneous calibrated optical spectra from the NOT, we show in Fig. 8 the spectra from
Days 25 and 46 that  flank the STIS epoch.  The profile was weaker and blended on day 25 but by day 46 had fully
developed the same profile as the [O I] 6300\AA\  and related lines.  
The higher ionization stages of N at this time were not observed in the UV and completely dominated by strong metallic absorption lines.  It is possible, however, that emission was already  beginning to develop in the N III] 1750\AA\ and N IV] 1486\AA\ lines.

On day 99, during the transition stage, strong emission on  three carbon (C$^+$-C$^{+3}$) and four nitrogen (N$^+$ - N$^{+4}$) ions were visible.  The profiles were very asymmetric, with the blueshifted peak at $\sim$-800 km/s dominating the line profiles, and  with little or no redshifted counterpart.    The +800 km/s peak was accentuated in the N IV] and N V lines, the red to blue component ratio increasing with ionization stage.  The emission wings were nearly identical in the ions of both species, extending to -1500 to -1800 km s$^{-1}$.   In addition, they displayed fine structure (better visible in [NII] and CIII] ), with individual peaks of width $\sim$200km s$^{-1}$ and distributed throughout the velocity range -1000$\leq$v$\leq$1000 km s$^{-1}$. None coincided with the velocities observed during the Fe curtain stage, and the peak at +600 km s$^{-1}$ observed during the opaque stage had disappeared or was blended with similar, lower velocity structures. The N IV] and C III] (isoelectronic) profiles 
were not quite identical, the emission in the interval -500 to 500 km s$^{-1}$ was
stronger for C III] than N IV], but the red to blue peak ratio was about the same.

\begin{figure}
   \centering
   \includegraphics[width=9cm]{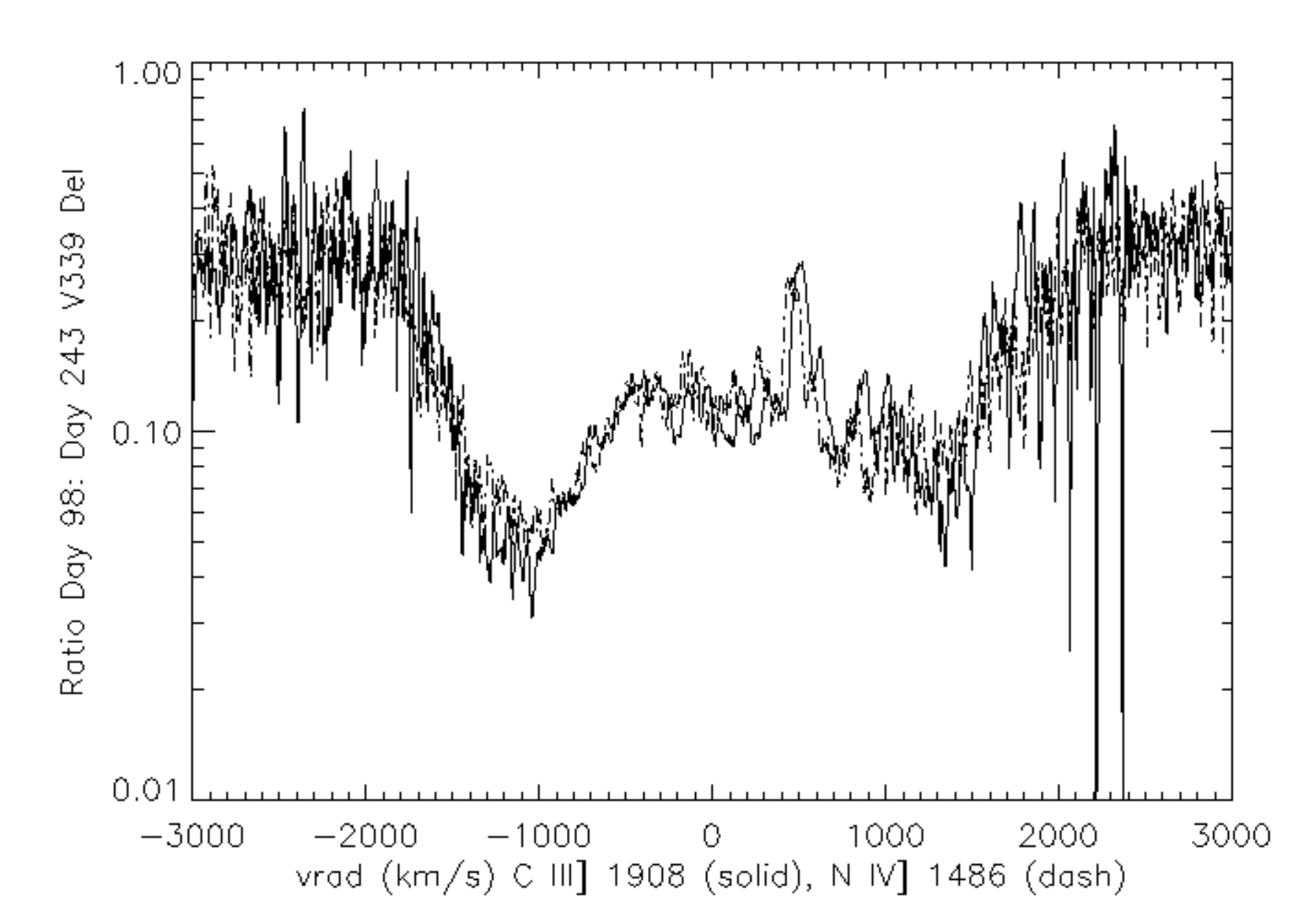}
      \caption{Ratio of the C III] 1909\AA\ flux on day 99 over day 248 flux (solid line) and same for the N IV] 1486\AA\ (dashed line).  Note the strong decrease in the strength of the +600 km s$^{-1}$ feature and the near identity of the variation for the two lines.  In both cases, the weaker component of the doublet is negligible.}
         \label{f6orig}
   \end{figure}

Our day 248 (UV) and 243 (optical) spectra again showed similar line profiles in term of substructures in the velocity space, but
with significantly changed relative intensities. The blue peak at $\sim$-800 to -900 km s$^{-1}$ faded faster than any other
substructure within the line profile, contrarily to the peak at +600 km s$^{-1}$ which faded at a slower pace (Fig.\ref{f6orig}). 

\subsection{Extinction}

The extremely rich interstellar absorption spectrum toward V339 Del will be treated in a separate paper.  We have,
however, derived independent extinction constraints based on comparing the optical and ultraviolet line profiles  to the
LAB 21 cm H I profile (Kalberla et al. 2005) in the velocity range -100 $\le$v$_{LSR}$ (km s$^{-1}$)$\le$ 100.   We show (Fig. 10) the comparison of the 21 cm line with tracers of the neutral gas,  O I 1304\AA\  (day 28) and NaI D (Day 99).  In the velocity range
covered by both lines, the integrated neutral hydrogen column density, $N_H$ was $\sim 1.1\times 10^{21}$cm$^{-2}$. This is consistent with the minimum H I absorption derived from the {\it Swift} spectra obtained during the SSS phase.  The extinction derived from the this N$_H$ is E(B-V)$\approx$0.17 (G\"uver \& \"Ozel 2009), although we cannot correct for the molecular fraction so this refers to only the atomic component.  The value is consistent with that reported by  Munari et al. (2015), E(B-V)=0.18, based on optical spectral calibrators.

   \begin{figure}
   \centering
   \includegraphics[width=9cm]{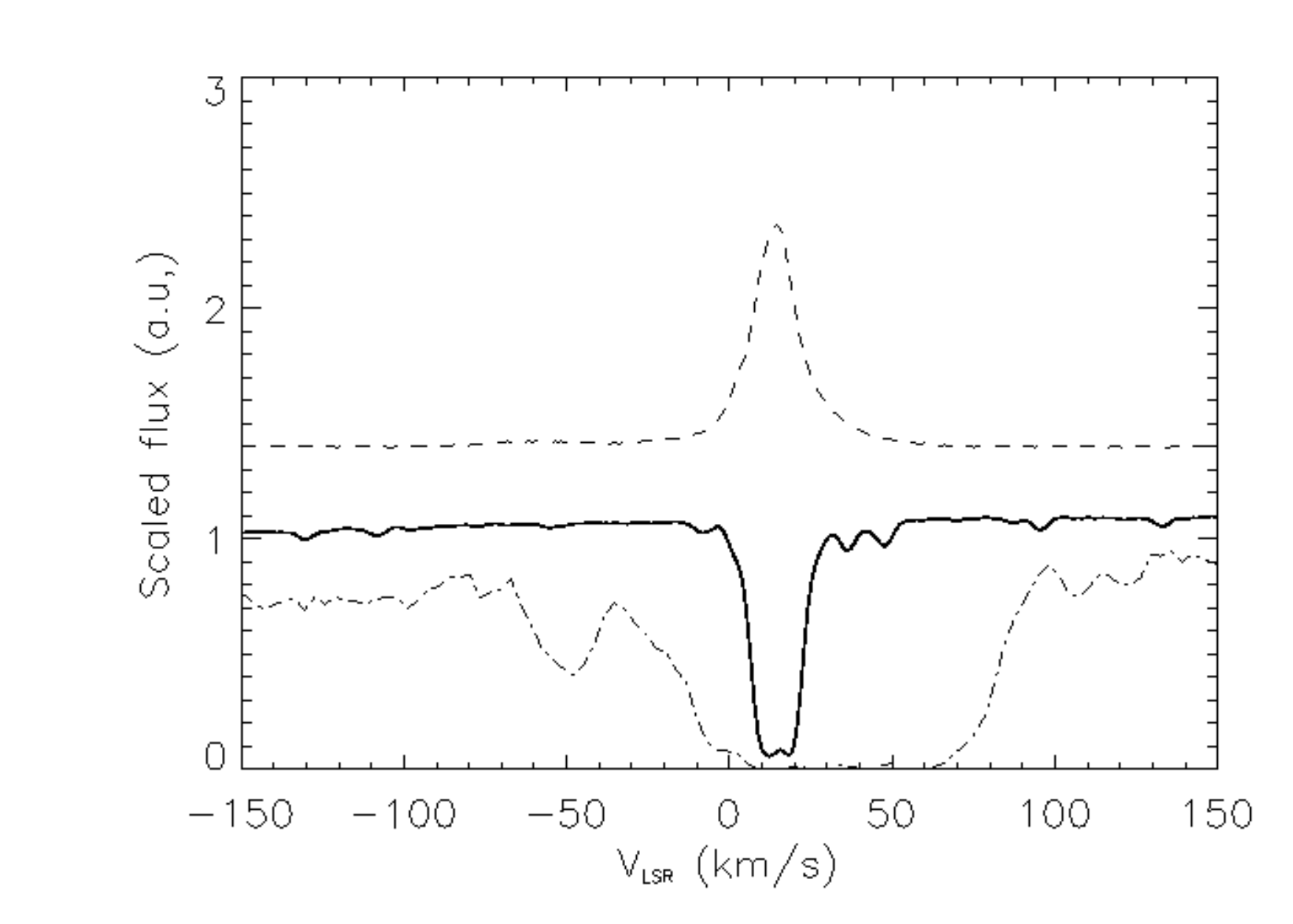}
         \caption{Interstellar line profiles for O I 1303\AA\  (dot-dash) and Na I D1 (solid)  compared with the LAB 21 cm H I
profile (dash) in LSR velocity.  The O I profile is from the Day 99 STIS spectrum, that for Na I D1 is from the Day 14 FIES
spectrum. }
         \label{f9}
   \end{figure} 

   \begin{figure*}
   \centering
   \includegraphics[width=14cm]{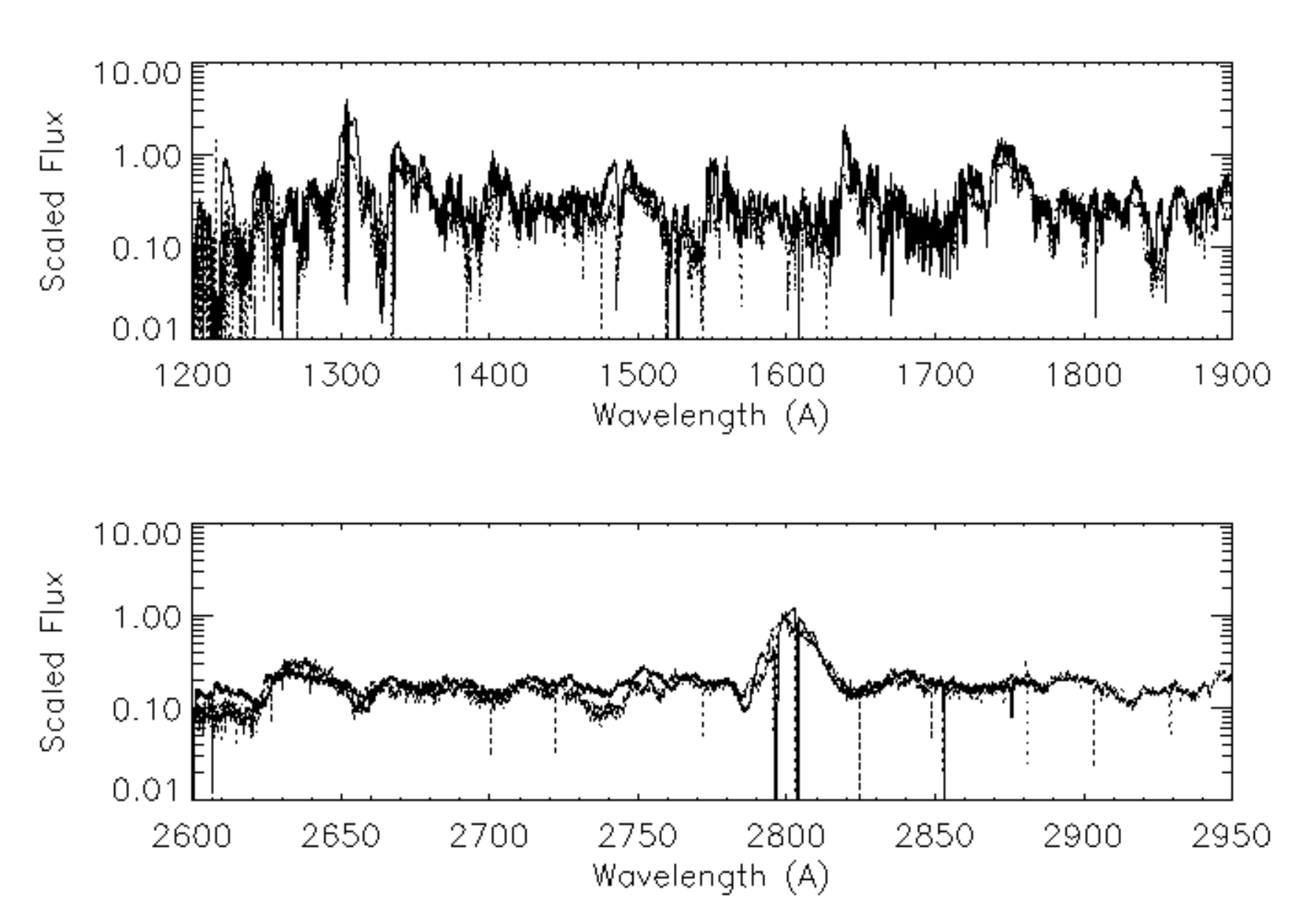}
      \caption{Comparison of V339 Del (solid, E(B-V)=0.18 and divided by 1.6) and OS And 1986 (E(B-V)=0.25) for Day
38.  The fluxes of each are scaled by $10^{-10}$ erg s$^{-1}$cm$^{-2}$\AA$^{-1}$. See text for discussion. }
         \label{f10}
   \end{figure*}

\subsection{Distance and luminosity}

Notably among recent novae, near infrared interferometric  observations  were obtained with  CHARA  (Schaefer  et  al.  
2014) almost immediately after discovery.  These showed an increase in  the angular size of the ejecta from day +1 (from  the
estimated outburst  onset) to day 27, from  which Schaefer  et  al.  determined a distance of  4.54$\pm$0.59 pc.  They assumed 
an expansion velocity of 650 km s$^{-1}$,  based on the Si II 6347, 6371\AA\ doublet.  Since these are intrinsically weaker lines than the Balmer or any of the usual optically thick tracers of the ejecta, they form closest to the same pseudo-photosphere that produces both the optical and infrared continuum during the earliest stages of the expansion.  

To obtain independent values for the extinction and distance, we used a template nova, the classical CO nova OS And 1986.  Our method was as follows.  OS And was observed with {\it IUE} at high resolution (R$\approx$10000) during the Fe curtain stage,  from day 4 to day 22  after discovery, after which observations were interrupted by
the outburst of SN 1987A (Schwarz et al. 1997).  This was the only bright CO nova observed in the ultraviolet at a stage comparable to that of the first NOT/STIS sequence.  We compared our day 35  STIS  spectra with high resolution IUE
spectra of  OS And  obtained between day 12 (LWP 9717, 9719) and day 22 (SWP 29981)  after optical
discovery (1986 Dec. 17 and Dec 27 respectively) using E(B-V)=0.18 for V339 Del and the extinction for OS And obtained by Schwarz et al.(1997), E(B-V)=0.25$\pm$0.05.  This provided a close match, with the flux for V339 Del being about 1.6$\pm$0.1 times higher than OS And (1200-1900\AA).  The long wavelength {\it IUE}  exposures were slightly earlier and at this stage that can make a difference, but they were similar.   The comparison is shown in  Fig. 11.   The  Fe curtain features correspond well after scaling, especially near 1750\AA, the part of the extinction curve that is nearly flat.  Thus, for a distance of OS  And of 5.1$ \pm$ 1.5 kpc (Schwarz et al. 1997), the derived distance to V339 Del is about 4.1 kpc.  This is consistent with the CHARA near infrared  interferometric determination.

During the Fe curtain  phase, before the detection of soft X-rays, the high optical depth of the ejecta insures that the
 irradiation from the central source is completely reprocessed into the visible and ultraviolet.  Gehrz et al. (2015) have published near infrared (JHK) photometry that is contemporaneous with the NOT/STIS observations during the first 100 days   Since we do not have simultaneous infrared fluxes,  any bolometric luminosity estimate is a lower limit but,  given the continuum, it is likely that
only a small fraction of the flux is missing.   The 1200-7400\AA\ luminosity was $L_{bol} \ge (5\pm 0.5) 10^{4}$L$_\odot$. 
Even as a lower limit, this is at or above the Eddington luminosity for a Chandrasekhar mass white dwarf.  Since, as we
have argued, the ejecta were neither spherical nor completely covering, this estimated luminosity is an even more restrictive lower bound.  Note that this $L_{bol}$  does not refer to the {\it maximum} luminosity during the explosion, but to the state of the WD long after the ejecta were in free expansion.  

\subsection{Geometry}

To obtain a picture of the geometry of the ejecta, since the solid angle is needed for obtaining its mass, we used
Monte Carlo modeling, as in  our recent papers on T Pyx (De Gennaro Aquino et al. 2014), V959 Mon (Shore et al. 2013),
and our companion study of V339 Del (Shore et al. 2014).  Assuming a constant mass for the ejecta and an inverse cubic
radial dependence for the density obtained from a linear velocity law, we set the maximum expansion velocity to 2500 km s$^{-1}$, guided by the highest velocities observed on the UV resonance lines and the optical Balmer lines, and explored a range of inclinations ($i$), ejecta opening angles ($\theta_o$, the outer angle and $\theta_i$, the inner angle), and fractional inner radius, $\Delta R$.  The ionization was not explicitly modeled from first principles, but a comparison was made between the individual line profiles for different ions and a range of models.   We show sample results for the low and high ionization stage lines compared with the observed profiles in Fig. 12.  The asymmetries in the profile are yet to be understood, but the emission lines in all novae display them when they are optically thin.


With the inner angle fixed at 90$^o$ and $v_{max} = 2500$ km s$^{-1}$, the range of allowable parameters is $35^o \le
i \le 55^o$, $10^o \le \theta_i \le 30^o$ and $\Delta R/R \approx 0.3$.  As shown in Fig. 12, the profiles are
reproduced by the bipolar geometry.  Individual features are, however, random (this procedure is based on a Monte Carlo
simulation), so even the relatively large dissimilarities in the peaks can be explained as chance asymmetries of the
ejecta,.  The {\it essential} feature of the profiles is that they require a bipolar structure and that the peaks are
merely projection effects.    Regarding
the geometry, with higher quality spectra covering longer times, it is now emerging that the non-spherical structure -- that is roughly bipolar -- is generic (see also Woudt et al. 2009).  This has been emphasized by Shore (2013, 2014), and Ribeiro et al. (2013a,b, 2014).  The ejecta are, however,  not really axisymmetric.  Although asymmetries have been noted for some time, they are based on  early spectra when it could be attributed to optical depth effects (see, e.g., Shore 2012).  In our case, the profiles are optically thin and the asymmetries are intrinsic.  The analysis presented here shows that the ejecta are far from ideal structures, and the substructures seem to have a random distribution.

   \begin{figure}
   \centering
   \includegraphics[width=9cm]{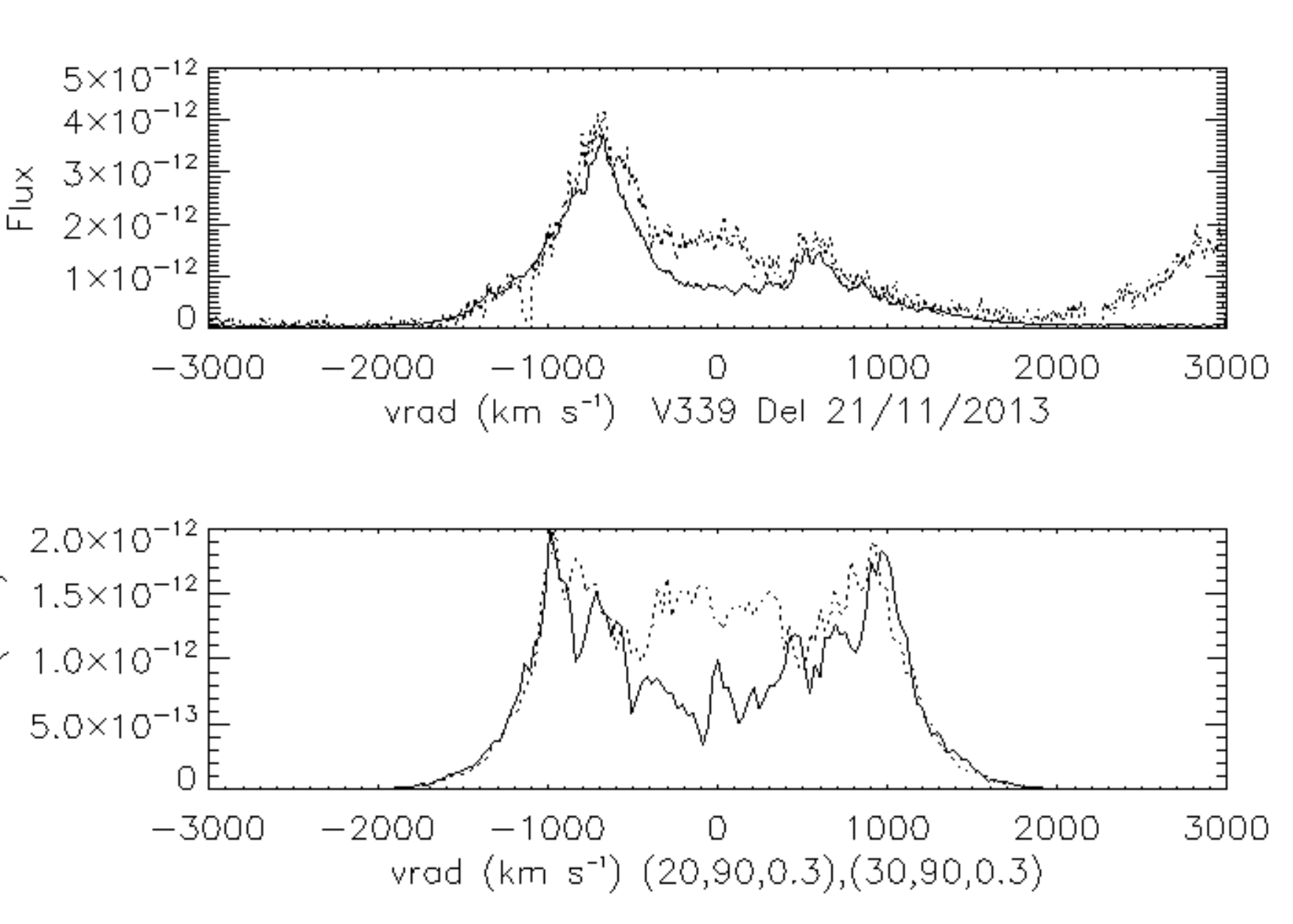}
      \caption{Sample model profiles for the Day 99 spectra.  Top: N IV] 1486\AA (solid), He II 1640\AA (dot).  Bottom:
model profiles (flux in arbitrary units).  The model parameters are listed for the two simulations. DA RIFARE.}
         \label{f11}
   \end{figure} 
   
   
   The different ionization stages can be reproduced with the bipolar ejecta  with a single
parameter change, the outer cone angle.  Keeping all other parameters fixed, the profiles of the higher ions require a
narrower region of line formation (for {N II] as for the other lower ionization stages the outer angle was 10-20$^o$
while for the higher ions it was closer to 30-40$^o$.


\subsection{Electron density}

The [O III] 4363, 4959, 5007\AA\ lines were used to estimate the electron density in the velocity interval of highest
S/N ratio, in the range from -1000 to 1000 km s$^{-1}$, for four epochs (days 99, 122, 243, and 435) for which we have
calibrated NOT spectra and when the ejecta displayed the nebular lines.  It was never possible to perform a similar
analysis for the [N II] lines because, during this epoch, the 6548, 6583\AA\  doublet was always too severely blended
with H$\alpha$.   The adopted reddening was E(B-V)=0.18 (Munari et al. 2015) and we assumed an electron temperature of
$10^4$K.  The latter does not critically affect  the result (Osterbrock \& Ferland 2005).  Two related properties result from the analysis.  For 
 ballistic ejecta, the electron density varies as $n_e \sim r^{-3}$, and the radial position
is exchangeable with the radial velocity once the matter is transparent (i.e. the velocity is linearly dependent on radius).   Thus, regardless of the symmetry of the ejecta, 
the same electron density applies to any transition within the same velocity interval.  The H$\beta$ line was used to decompose the blend between H$\gamma$ and [O III] 4363\AA, after having confirmed that the line profile for the Balmer lines was the same as the other optically thin transitions.  For the first observation, day 99, this is not certain since there was some residual self-absorption for the hydrogen
transitions.  But thereafter, the hydrogen and [O III] (and [N II] 5755\AA) profiles agree in detail.   The narrow emission features  have widths of order 50 - 100 km s$^{-1}$, except for the strongest at $\pm$550 km s$^{-1}$ that are of order 200 km s$^{-1}$, and the wings extend to $\pm$1500 km s$^{-1}$.  

The ratio $[F_{4959}+ F_{5007}](\Delta v_{rad},t)/F_{4363}(\Delta v_{rad},t)$ then provides the $n_e(t)$. This result is
shown in Fig.\ref{f12} and listed in Table \ref{nef}.    The uncertainties in n$_e$ were derived from the noise in the intensity ratio within the specified velocity interval.   For comparison, we also plot the expected temporal variation in the density, $n_e \sim
t^{-3}$;  a least squares fit to a power law gave the exponent -2.4, including the marginally nebular spectrum from day 99.  An
additional check that the ejecta were in the nebular stage is that F(5007)/F(4959)$\approx$3; this was found at every point in the line profile  after day 99.  As a comparison, in Fig.\ref{f12} we also show the time dependent H$\beta$ fluxes.   Since the Balmer lines are formed by recombination once the ejecta turn transparent in the UV, H$\beta$ should display the same variation as the
independently determined $n_e$; the line in Fig. 13 is the t$^{-3}$ variation.    Sample fluxes are given in Table 3 (a more complete analysis will be presented along with the abundances in the next paper in this series).  

\begin{table}
\centering
\caption{The derived physical properties}
\label{nef}
\begin{tabular}{lccc}
\hline
Day & n$_e$ (cm$^{-3}$) & filling\ factor  & Mass ($10^{-5}$M$_\odot$)\\
\hline
98  & $(1.2\pm0.4)\times 10^7$& 0.7$\pm$0.4 & 3.7$\pm$2.0\\
118 & $(1.7\pm 0.5)\times 10^7$ & 0.2$\pm$0.1 & 3.0$\pm$1.7\\
243 & $(4.5\pm 1.0)\times 10^6$ & 0.07$\pm$0.03 & 2.1$\pm$1.2\\
435 & $(4\pm 1)\times 10^5$  & 0.1$\pm$0.05 & 2.1$\pm$1.2 \\
\hline
\end{tabular}
\end{table}

   \begin{figure}
   \centering
   \includegraphics[width=8cm]{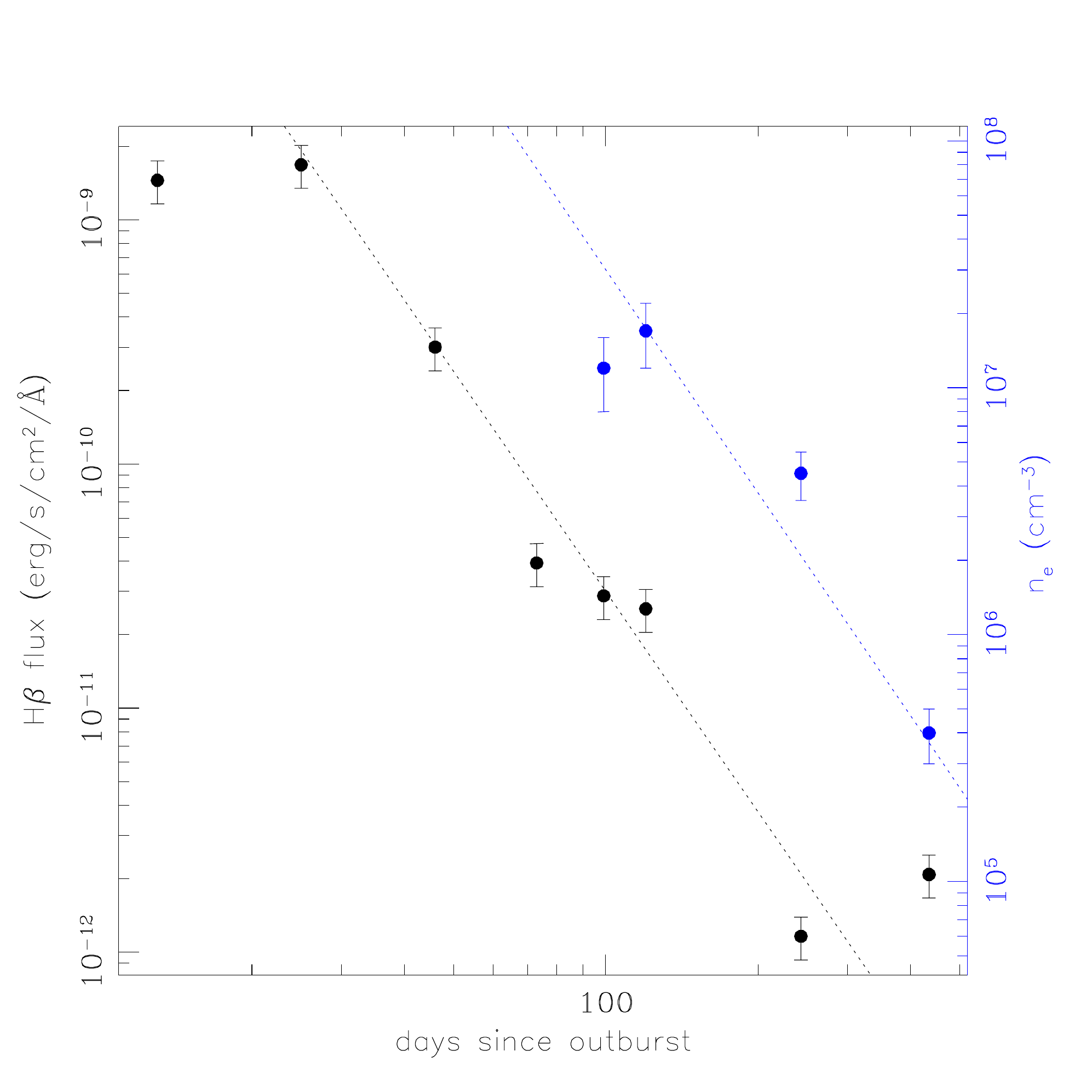}
      \caption{H$\beta$ flux (black) and electron density time dependence. The H$\beta$ integrated flux has been
measured from the full set of NOT/FIES spectra. The electron density has been obtained from the [O III] analysis, as 
described in the text only for the NOT/FIES spectra in nebular conditions. The error bars are 1$\sigma$ errors. The
two dotted lines indicate the $t^{-3}$ dependence and are {\it not a fit to the data points}. }
         \label{f12}
   \end{figure}

The X-ray source was active during most of the interval described in this paper.  Hence we expect that the radiative balance in the optically thin ejecta were still governed by photoionization processes.  The densities were not yet so low that the ejecta had reached the frozen-in
state, which occurred after Day 248 (see, e.g., Shore (2012) for discussion).  We will discuss this stage in a future paper.


\begin{table*}[h]
\caption{Sample line integrated flux from NOT/FIES set of spectra. Note that the fluxes are in erg s$^{-1}$cm$^{-2}$ and no
reddening correction has been applied. Estimated errors are 20\% systematic and 10\% random.  Those transitions that were especially weak are marked with a colon.}
\label{}
\begin{tabular}{lccccccc}
\hline\hline
Day & H$\beta$ & Fe II 5169 & [O I] 6300 & [Fe VII] 6087 & He II 4686 & [O I] 5577 & [N II] 5755 \\
\hline
13  & 1.45E-9: & 1.91E-10 & 9.68E-11 & ...     & ...      & 6.19E-11 & ... \\
24  & 1.68E-9  & 1.37E-10 & 1.41E-10 & ...     & ...      & 7.25E-11 & 5.5E-11 \\
46  & 3.01E-10 & 1.73E-11 & 5.65E-11 & $<$6E-11& ...      & 3.71E-11 & ... \\
74  & 3.93E-11 & 1.79E-12 & 1.39E-11 & 3.1E-13 & 6.7E-12: & 4.22E-13 & 1.53E-11 \\
98  & 2.88E-11 & 1.43E-12 & 9.72E-12 & 3.2E-12 & 6.2E-12: & 2.15E-13 & 1.38E-11 \\
123 & 2.55E-11 & 1.41E-12 & 5.95E-12 & 3.3E-13 & 5.1E-12  & 1.76E-13 & 1.29E-11 \\
243 & 1.16E-12 & 2.1E-13  & 1.78E-12 & 5.0E-13 & 2.0E-12  & ...      & 1.58E-11 \\
435 & 2.08E-12 & ...      & 6.1E-13  & 7.2E-14 & 3.5E-13  & ...      & 2.86E-12 \\
\hline
\end{tabular}
\end{table*}


\subsection{Filling factor and mass}

The filling factor, $f$, was computed using the (integrated) H$\beta$ luminosity from each NOT spectrum and $n_e$
obtained from the [O III] analysis, following the method described in Shore et al. (2012).  One difference here is the
number of optical spectra obtained during the later stages of the expansion, when the ejecta  were optically thin. 
It was, therefore, possible to see if $f$ varied with time.  The overall ejecta geometry was obtained from the model line profiles.    The ejecta were optically thin in the Lyman transitions.  The results for $f$ are shown in Fig.\ref{f13}, together with the derived ejecta mass, and listed in Table 2.  The filling factor and mass were roughly constant after Day99, $f \approx 0.07-0.2$
and $M_{ej} \approx (2-3)\times 10^{-5}$M$_\odot$.   The [N II] 6548, 6583\AA\ doublet remained too blended with
H$\alpha$ during the period discussed here for an  independent determination of $n_e$.  As with the electron density, the ejecta were 
still in the transition to nebular on Day 99, during which time the Lyman series was still optically thick, hence the
overestimate of the filling factor on that date.    The contrast between the diffuse gas and the filaments was, however, reduced in the still relatively compact ejecta, and became more distinct in the later spectra.

   \begin{figure}
   \centering
   \includegraphics[width=8cm]{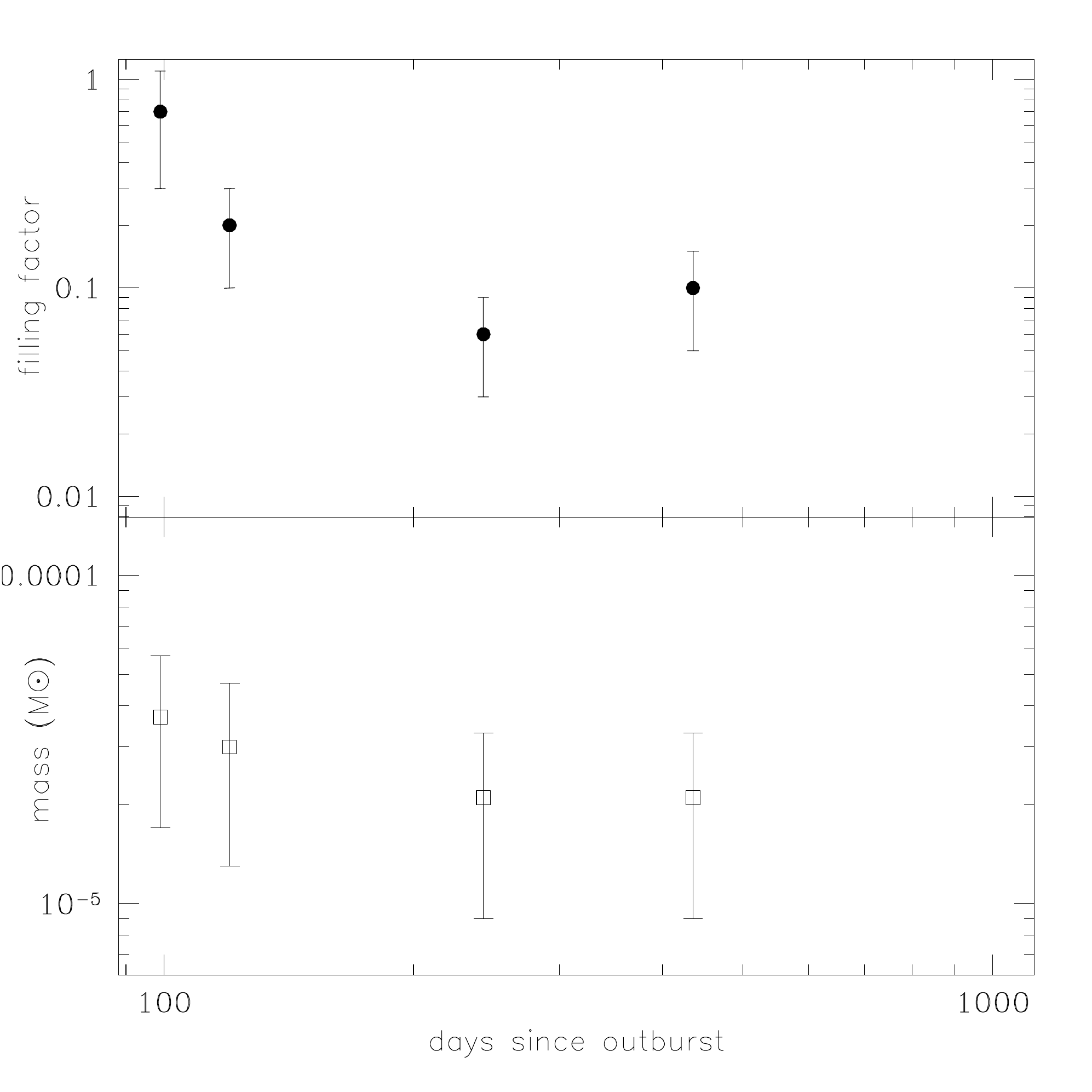}
      \caption{Time dependence of the inferred filling factor and ejecta mass.  Error bars are 1$\sigma$.   See text for
discussion. }
         \label{f13}
   \end{figure}
   
The derived  mass, and filling factor for V339 Del are similar to the other recently observed classical nova, V959 Mon
(Shore et al. 2013) but  higher (for both) than T Pyx by the same analysis (Shore et al. 2014).    That the filling factors are relatively small also supports the idea that there is a  physical variance in the line profiles that is well-captured by the Monte Carlo modeling.   The mass is in line with predictions of TNR modelling (see, e.g., Starrfield et al. 2012).  The higher masses previously reported in analysis of classical novae have not included this information and, as noted in Shore (2013) and Shore et al. (2014), have likely systematically over-estimated the ejecta masses.   

\section{Conclusions}

In its basic properties,  V339 Del is not remarkable for a CO nova.  For the quantity, quality and timing of the multiwavelength observations, and the enormous spectroscopic archive amassed, however, it surpasses even the optical spectroscopic database for the  classical nova DQ Her (Stratton \& Manning 1939; Payne-Gaposchkin 1957).   Our aim was to systematically sample the range of spectroscopic diagnostics at critical stages during the evolution of the outburst.  By combining X-ray, ultraviolet, and optical spectroscopy and spectrophotometry, we have obtained a set of benchmark parameters for this nova.

Abundances for key elements were not derived at this stage.  We intend to perform a complete analysis once the SSS phase has definitively ended and the ejecta are in freeze-out  (e.g., Vanlandingham et al. 2001).  The geometry and relatively low filling factor of the expanding matter, and the requirement to treat  time dependent excitation and ionization balances because of the wide range in densities, makes a photoionization analysis too preliminary at this time.  The next paper in the series will address this question.  The {\it HST/STIS}  observations indicate that  there was no substantial Ne overabundance.  In all spectra following the nebular phase transition, neither  [Ne IV] 1602\AA\ nor [Ne V] 1575\AA\  were detected above the noise and were an order of magnitude weaker than any of the previously studied ONe novae (reported in De Gennaro Aquino et al. 2014).  The prominence of the C and N lines throughout the optically thick stage, including the infrared (ATel 5336, 5337), point to the 
overabundances of the light elements.   It should be noted that no CO detections were reported during the early expansion.


Our main conclusions are:

\begin{itemize}
\item The maximum expansion velocity was $\sim$ 2500 km s$^{-1}$ measured from the emission line wings.  During the early Fe curtain stage, all transitions that were strongly coupled to ultraviolet resonance and low lying states displayed P Cyg absorption troughs whose profiles were consistent with ejecta of fixed mass in ballistic expansion.   Specifically, the velocity field was consistent with a linear radial expansion law and there was no evidence of secondary ejections or changes in the maximum velocity of the lines over time.  There was no evidence for a wind during the later stages 
of the outburst, and there is no indication of one during the Fe curtain phase. 
\item The ejecta were represented by a global biconical geometry.  There were, however, distinct asymmetries between the lobes that persisted and became more distinct as the higher velocity matter turned transparent.  The explosion, as now appears to be a more general result, was not axisymmetric.  Individual knots were present that span several hundreds of km s$^{-1}$ and correspond to tens of percent of the radial extent of the ejecta for ballistic expansion.     The ejecta were highly fragmented over a large range of velocity scales (hence radial sizes) but organized in a comparatively small number of distinct structures seen during the P Cyg absorption stage.   The emission line profiles all displayed extremely narrow structures, which was duplicated in virtually all ionization stages. 
\item During the Fe curtain maximum, the luminosity in the 1150-7500\AA\ band -- that should contain the bulk of the emission -- was {\it at least} $5\times 10^4$L$\odot$.  In light of the conclusions about the ejecta geometry, the actual luminosity at peak could have been significantly higher.  To date, the {\it bolometrically} most luminous CO nova was LMC 1991, which was {\it measured} above $10^5$L$\odot$, and V339 Del might have reached similar values (depending on the flux fraction lost from our line of sight).
\item A distance of $\sim$4.2 kpc was obtained using a comparison with {\it IUE} high resolution spectrophotometry of the CO nova OS And 1986.   The extinction was derived from the Ly$\alpha$ profile and from comparisons between the velocity extent of the optical  Na I D and UV interstellar resonance lines and the 21 cm LAB spectrum along the line of sight.   The derived value was 0.2 $\pm$0.04, consistent with published values based on early optical observations.  This implies a minimum neutral hydrogen column density of 1.2$\times 10^{21}$ cm$^{-2}$ that is consistent with the last fit value from {\it Swift} X-ray spectrophotometry during the late SSS phase.
\item The recombination-dominated transitions, such as He II 1640\AA, 4686\AA, and the Balmer lines, followed the same temporal trend as the electron density, varying as $t^{-3}$ during the nebular stages.  
\item The filling factor was between 0.1 and 0.3.  The ejecta were consistent with chemical homogeneity, but this conclusion is tentative pending a full abundance analysis.
\item The derived mass of the ejecta was (2-3)$\times 10^{-5}$M$_\odot$.
\end{itemize}

Our program of observations is continuing into the late nebular stage and will be treated in the next paper in this series.

   \begin{figure}
   \centering
   \includegraphics[width=8cm]{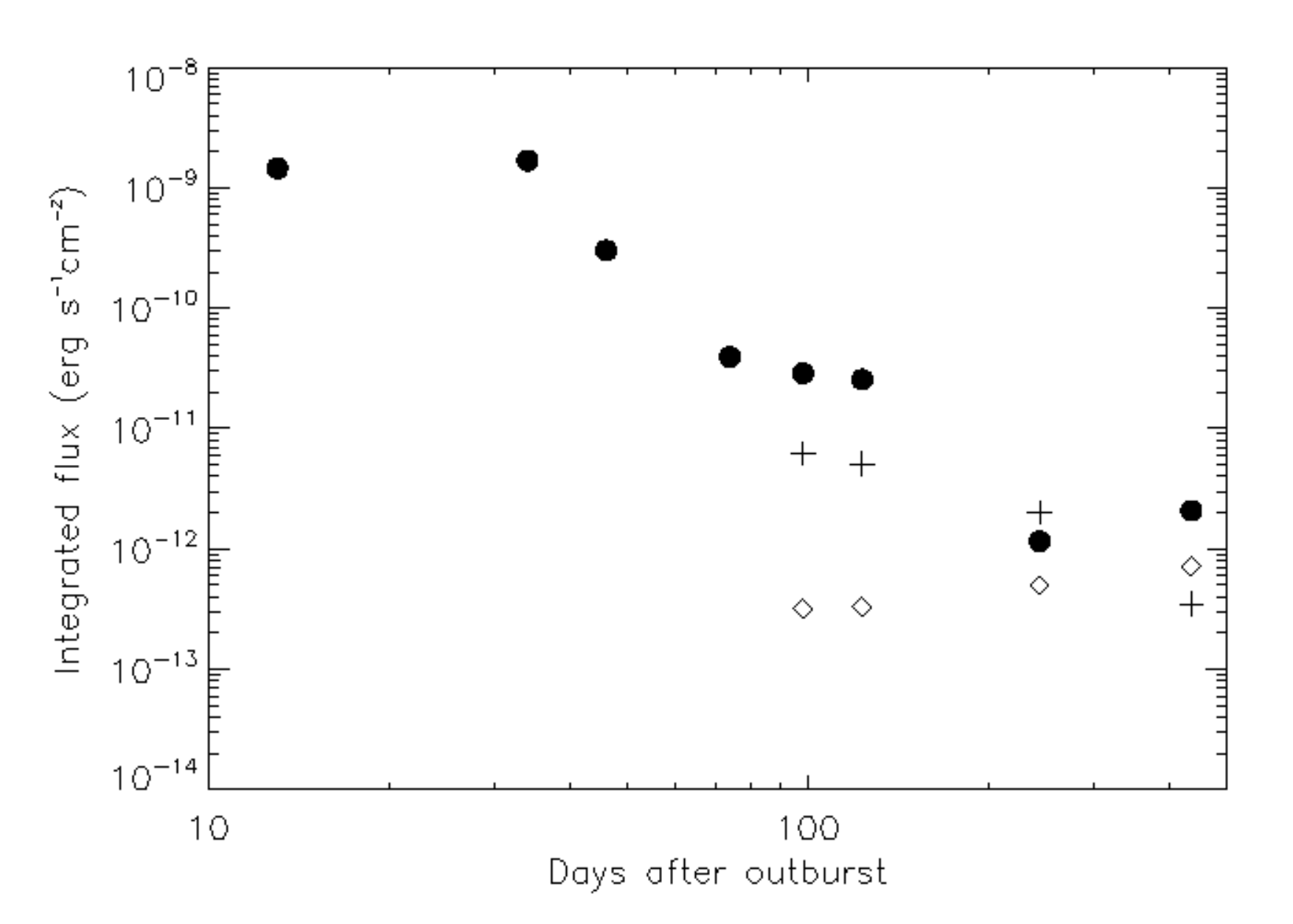}
      \caption{Flux variations for select lines from the NOT sequence.  Dot: H$\beta$, plus: He II 4686\AA; diamond: [Fe
VII] 6087\AA. }
                      \label{}
   \end{figure}

\begin{appendix}
\section{1200-7300\AA\ full spectral distributions for days 35 (Fe curtain), 99 (transition stage), and 248 (nebular stage)}
\begin{figure*}[h!]
   \centering
   \includegraphics[width=13cm,angle=0]{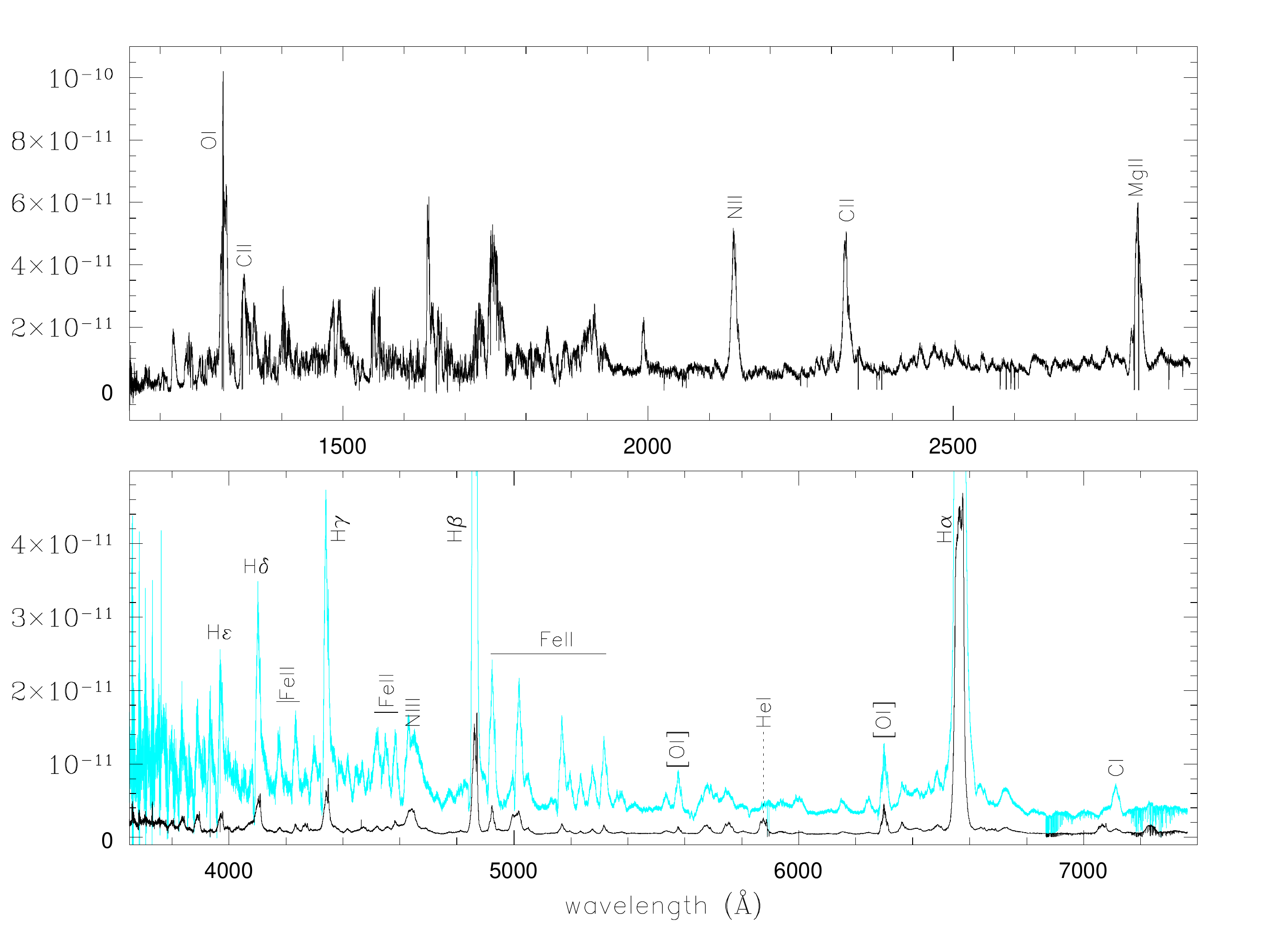}
      \caption{Panchromatic spectrum of V339 Del on Day 35 obtained combining the STIS (day 35) and FIES (day 25 in
cyan and day 46 in black) observations. No extinction corrections have been applied.}
                      \label{A1}
   \end{figure*}

   \begin{figure*}[b]
   \centering
   \includegraphics[width=13cm,angle=0]{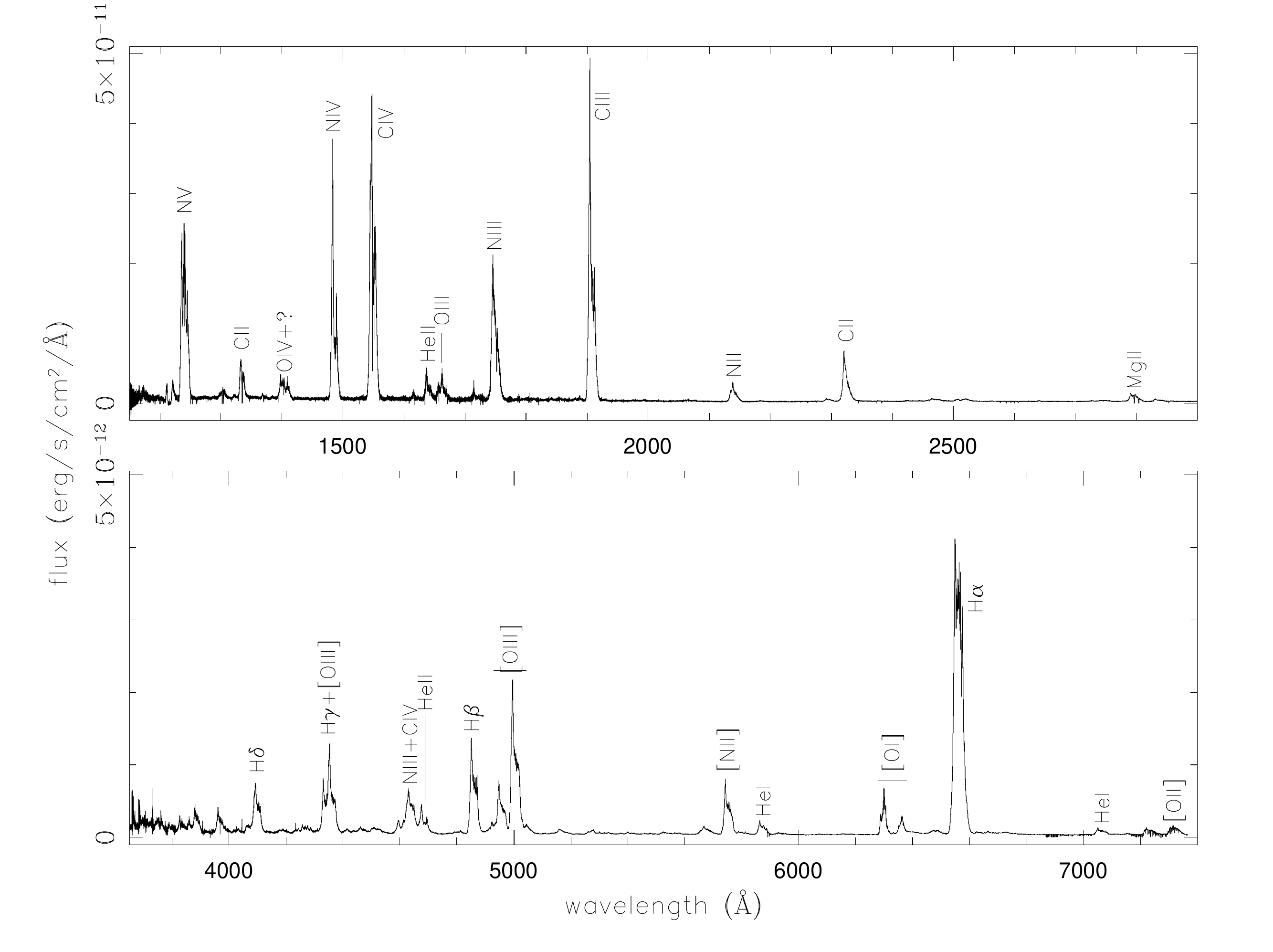}
      \caption{Panchromatic (STIS+FIES) spectrum for Day 99.}
                      \label{A2}
   \end{figure*}

   
   \begin{figure*}[ht]
   \centering
   \includegraphics[width=13cm,angle=0]{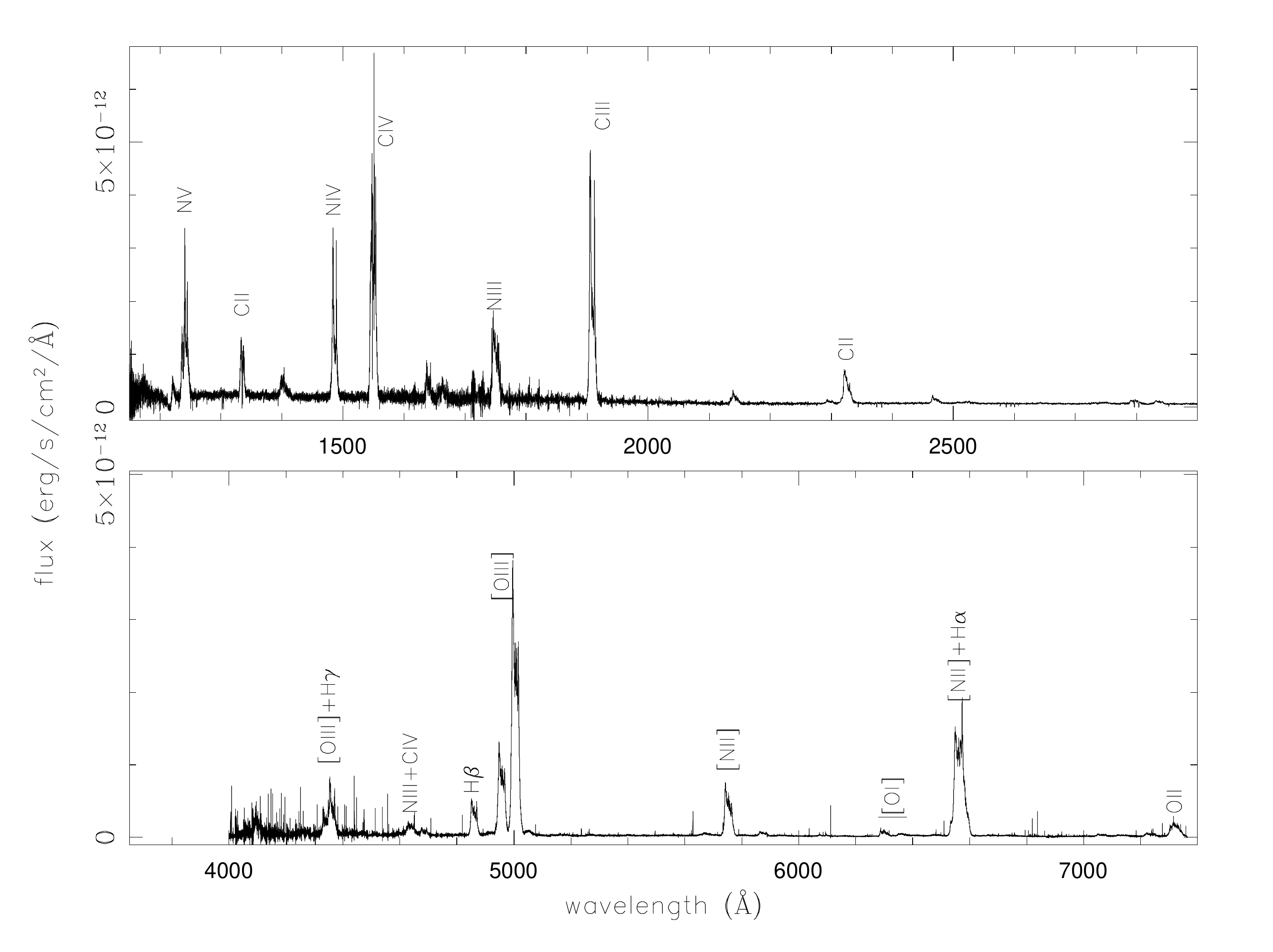}
      \caption{Same as Fig. A2 for Day 248.}
                      \label{A3}
   \end{figure*}
 
\end{appendix}

\begin{acknowledgements}

Based on observations made with the Nordic Optical Telescope, operated by the Nordic Optical Telescope Scientific Association at the Observatorio del Roque de los Muchachos, La Palma, Spain, of the Instituto de Astrofisica de Canarias. The research leading to these results has received funding from the European Union Seventh Framework Programme (FP7/2007-2013) under grant agreement No. 312430 (OPTICON).  Sumner Starrfield acknowledges partial support from NASA and NSF grants to ASU.  CEW acknowledges partial support from NASA Swift grant NNX14AC54G.  Simone Scaringi acknowledges funding from the Alexander von Humboldt Foundation.   

Based on observations made with the NASA/ESA Hubble Space Telescope, obtained [from the Data Archive] at the Space Telescope Science Institute, which is operated by the Association of Universities for Research in Astronomy, Inc., under NASA contract NAS 5-26555. These observations are associated with program \# 13828.   Support for program \#13828 was provided by NASA through a grant from the Space Telescope Science Institute, which is operated by the Association of Universities for Research in Astronomy, Inc., under NASA contract NAS 5-26555.

We thank our collaborators in the ARAS group for their remarkable diligence, skill, and persistence in obtaining spectroscopy of this and other novae  and for the resulting archival treasures.    The program described here was planned during a workshop in Pisa in July 2013, fortuitously  just before the announcement of V339 Del, and we thank the participants or their insights, especially A. Caleo, C-C. Cheung, J. Jos\'e, J-U Ness, and B. Warner.  We also thank  D. Gies, P. Hauschildt,  D. Korcakova, P. Kuin, P. Selvelli,   F. Walter, and P. Woudt for discussions and exchanges.
\end{acknowledgements}



\end{document}